\documentclass[prx,a4paper,reprint,showpacs]{revtex4-1}
\usepackage{color}
\usepackage{graphicx}                          
\usepackage{amssymb,amsfonts,amsmath}
\usepackage[utf8]{inputenc}
\usepackage{bbm}
\usepackage{pict2e}
\usepackage[export]{adjustbox}

\begin{document}
\title{How Damage Diversification Can Reduce Systemic Risk} 
\author{Rebekka Burkholz} 
\email{rburkholz@ethz.ch}
\affiliation{ETH Zurich, Chair of Systems Design \\Weinbergstrasse 56/58, 8092 Zurich, Switzerland}
\author{Antonios Garas} 
\email{agaras@ethz.ch}
\affiliation{ETH Zurich, Chair of Systems Design \\Weinbergstrasse 56/58, 8092 Zurich, Switzerland}
\author{Frank Schweitzer}
\email{fschweitzer@ethz.ch}
\affiliation{ETH Zurich, Chair of Systems Design \\Weinbergstrasse 56/58, 8092 Zurich, Switzerland}

\begin{abstract}
We consider the problem of risk diversification in complex networks. Nodes represent e.g. financial actors, whereas weighted links represent e.g. financial obligations (credits/debts).
Each node has a risk to fail because of losses resulting from defaulting neighbors, which may lead to large failure cascades. 
Classical risk diversification strategies usually neglect network effects and therefore suggest that risk can be reduced if possible losses (i.e., exposures) are split among many neighbors (exposure diversification, ED). 
But from a complex networks perspective diversification implies higher connectivity of the system as a whole which can also lead to increasing failure risk of a node.
To cope with this, we propose a different strategy (damage diversification, DD), i.e. the diversification of losses that are imposed on neighboring nodes as opposed to losses incurred by the node itself. 
Here, we quantify the potential of DD to reduce systemic risk in comparison to ED.   
For this, we develop a branching process approximation that we  generalize to weighted networks with (almost) arbitrary degree and weight distributions. 
This allows us to identify systemically relevant nodes in a network even if their directed weights differ strongly.
On the macro level, we provide an analytical expression for the average cascade size, to quantify systemic risk. 
Furthermore, on the meso level we calculate failure probabilities of nodes conditional on their system relevance. 

\end{abstract}

\maketitle

\section{Introduction}
\label{sec:intro}

In the course of globalization and technical advancement, systems become more interconnected and system components more dependent on the functioning of others \cite{Goldin2010, Stiglitz2002}. 
In particular for socio-economic networks \cite{Schweitzer2009c} and financial networks \cite{Haldane2011, Garas2008c} we observe an increase in coupling strength and complexity at the same time.
Examples are global supply chains, but also technical systems, as, e.g., power-grids in the USA and Europe \cite{Brummitt2012}.

Increased dependence, under normal conditions, has advantages for the efficient operation of systems. But it also makes systems more vulnerable if some of their components break down. Precisely, the failure of a few components can get amplified such that it  leads to system wide failure cascades.
Classical risk management theories suggest that higher risk diversification reduces the failure risk of the component and consequently the one of the system as a whole. But these conclusions are based on the assumption that a component only depends on other components that are independent among each other. 

This assumption does not hold  in most real world systems.
Therefore, we model such dependences by a network where system components are represented as nodes and direct interactions as links between nodes.
Even if the neighbors of a node are not linked directly, they can be coupled indirectly through the network. 
This effect increases if nodes have a high degree, i.e. are well diversified in a financial context. 

In order to study simple risk diversification strategies  in the context of systemic risk a model introduced by Watts~\cite{Watts2002SimpleModelof} has been adopted to financial  
networks of interbank lending \cite{Gai2010, Battiston2012a, Roukny2013}. 
We call this the \emph{exposure diversification} approach (ED), and in this paper we will contrast it with an approach where the \emph{impact} of a failing node is diversified, instead. 
In a banking network this would correspond to a policy where each financial institution is only allowed to get into a limited amount of debt. 
This \emph{damage diversification} approach (DD) has the potential to reduce systemic risk significantly, since it counterbalances the failure amplification caused by hubs. 
Such hubs, because of their large number of neighbors, considerably affect the network in case of failure.
Consequently, policy discussions center around the question of how to prevent the failure of hubs, e.g. by increasing their robustness (i.e. capital buffers in finance) \cite{Arinaminpathy2012, Roukny2013}.
Our findings from the DD approach suggest to complement such regulatory efforts by the mitigation of the \emph{impact} of the failures of well connected nodes.

In this paper we present simulations as well as analytic derivations of network ensemble averages in the limit of infinite network size, where two quantities on the system level are given: a) the degree distribution, which defines the number of direct neighbors of a node and thus limits their respective diversification strategies, and b) the distribution of robustness which is later defined by the failure threshold.
The analytic method (also known as heterogeneous mean field or branching process approximation) has been derived for the ED approach \cite{Gleeson2007}. 
But, it does not capture processes where the impact of a failing neighbor depends on its specific properties (e.g., its degree or robustness) as it is required for the treatment of the DD variant.
Therefore, we extend the branching process approximation to the latter case and generalize it for the application to weighted random network models, whose weight statistics could be deduced from data, taking also into account different properties of neighboring nodes. 
This way, we generalize the analytic treatment to match application scenarios, by better understanding the role of different risk diversification strategies. 
Our analysis provides an interesting mesoscopic perspective with the comparison of failure probabilities of nodes with different diversification strategies, which should accompany the study of macroscopic measures, such as the average cascade, since it is crucial for the identification of system-relevant nodes. 

\section{Modeling exposure versus damage diversification} 
In a weighted network with $N$ nodes a link with (non-zero) weight $w_{ji}$ between two nodes $j$ and $i$ represents an exposure of $i$ to its network neighbor $j$. Each node $j$ can fail either initially or later in (discrete) time $t$ because of a propagating cascading process. Its (binary) state then switches from $s_i(t) = 0$ (ok) to $s_i(t) = 1$ (failed), without the possibility to recover.

If the node $j$ fails, its neighbor $i$ faces the loss $w_{ji}$. The total amount of $i$'s losses sums up to its fragility
\begin{equation}
\phi_i(t+1) = \sum_{j} w_{ji} s_j(t).
\label{eq:fragility}
\end{equation}
If $\phi_i$ exceeds the threshold $\theta_i$ (i.e., $\phi_i \geq \theta_i$), then node $i$ fails as well. Hence, $\theta_i$ expresses the \emph{robustness} of node $i$.  In this way a cascade of failing nodes develops over time, which can even span the whole network.
 
We measure the cascade size by the final fraction of failed nodes
\begin{equation}
\rho_N = \lim_{t\rightarrow \infty}\frac{1}{N} \sum^N_{i=1} s_i(t),
\label{eq:risk}
\end{equation}
when no further failures are triggered. 
Any cascade stops after at most $N$ time steps, since at least one node needs to fail at a time to keep the process ongoing.

The cascade dynamics are fully deterministic for given thresholds and weights on a fixed network.
Still, many systems do not remain constant over time, and may also fluctuate by their exposure to large cascades. 
Consequently, it is reasonable to quantify the risk of large cascades with respect to macroscopic distributions that allow for microscopic variations of the weighted network and the thresholds. 
The average cascade size with respect to these distributions then defines our measure of systemic risk.

In such a setting, we study the influence of two diversification variants.
The difference between the ED and the DD approach is in defining the weights $w_{ij}$. 
Precisely, 
\begin{align}\label{eq:weights}
 w^{\rm{(ED)}}_{ij} = \frac{1}{k_j} \;;\quad
 w^{\rm{(DD)}}_{ij} = \frac{1}{k_i}
\end{align}
for \emph{exposure diversifications} (ED) and for \emph{damage diversifications} (DD) respectively. 
Here, $k_i$ denotes the degree of node $i$, i.e., the number of its neighbors. 
We note that in general $w^{\rm{(ED)}}_{ij} \neq w^{\rm{(ED)}}_{ji}$ and $w^{\rm{(DD)}}_{ij} \neq w^{\rm{(DD)}}_{ji}$. 
Still, we call the network undirected. 
This differs from the approach for directed networks \cite{Gai2010, Amini2010, Amini2012} where the neighbors whose failures impact a node are distinct from the ones who face a loss in case of the node's failure. 
In this case one node is exposed to the other, but not vice versa.
In our case, however, once there is a link between two nodes, each can impact the other, but the amount of the loss can be different.

In the ED case every neighbor is treated identical. I.e. a failure of any neighboring node $j$ exposes a node $i$ to the same risk of a loss ${1}/{k_i}$. 
The higher the degree $k_i$ of node $i$, the better it diversifies its total exposure of $\sum_{j \in \rm{nb}(i)} {1}/{k_i} = 1$, where $\mathrm{nb}(i)$ denotes the neighbors of node $i$. I.e., in the ED case, single failures of neighbors $j$ become less harmful to node $i$ if it has a larger number of neighbors.
On the other hand, the failure of a hub impacts many other nodes and is thus problematic from a system perspective.

In the DD case the impact of a hub is effectively reduced. 
I.e., the failure of a hub $j$ has total impact $\sum_{i \in \rm{nb}(j)} {1}/{k_j} = 1$, which reduces the impact on a neighboring node $i$ to $1/k_{j}$ instead of $1$ in the ED case.
Hence, better diversified nodes damage each of their neighbors effectively less in case of a failure.

We are interested in how the heterogeneity of such diversification strategies and the heterogeneity of thresholds impacts systemic risk for large systems. 
In both model variants the diversification strategies are determined by the degrees of the nodes.

Consequently, we study the fraction of failed nodes as an average over a whole class of networks characterized by a fixed degree distribution $p(k)$ and a fixed threshold distribution $F_{\Theta}(\theta)$ in the limit of infinitely large networks ($N \rightarrow \infty$).
The network generation method with fixed $p(k)$ is known as configuration model \cite{Molloy1995,Newman.Strogatz.ea2001Randomgraphswith}, where all possible network realizations (without multiple edges and self-loops, but with a prescribed degree sequence) are equally likely. 
The thresholds are then assigned to nodes independently of each other, and independently of their degree according to $F_{\Theta}(\theta)$, although the independence of the degree is not a necessary assumption.   

For the ED approach, the average fraction of failed nodes at the end of a cascade can be calculated on random networks with given degree distribution $p(k)$ and threshold distribution $F_{\Theta}(\theta)$ \cite{Gleeson2007}. 
To obtain the results, a branching process approximation was used, also known as heterogeneous mean field approximation (HMF) or as local tree approximation (LTA) \cite{Dodds2009}. 
This approximation was studied in many subsequent works.
It was generalized for directed and undirected weighted networks~\cite{Amini2010,Hurd2013}, it was shown to be accurate even for clustered networks with small mean inter-vertex distance~\cite{Melnik2011}, and the influence of degree-degree correlations has been investigated~\cite{Dodds2009,Payne2009}. 
According to a general framework introduced by Lorenz et al.~\cite{Lorenz2009b}, the ED and DD approach belong to the constant load class, where the ED is called the inward variant, while the DD is identified as outward variant. 
Still, the risk reduction potential of the latter has not been understood so far, since a system's exposure to systemic risk has been only explored on fully-connected \cite{Lorenz2009b} or regular \cite{Tessone2012} networks, where both model variants coincide. 

In order to study the DD approach on more general networks, we generalized and extended the existing approximations, which for the case of ED were proven to be exact \cite{Amini2010, Hurd2013}. 
Now, we can treat more general processes where the directed weights in an undirected network can depend on properties of both nodes, the failing as well as the loss facing one. Here, in contrast to~\cite{Amini2010}, nodes can depend on each other, and in contrast to~\cite{Hurd2013} they do so in an non-symmetric way. 

We show in Section \ref{sec:results} that our approach leads to very good agreements with simulations on finite Erd\"os-R\'enyi networks and can also be applied to a more realistic setting where, e.g., the nodes' degrees follow a scale free distribution.

\section{Analytic framework}\subsection{Local tree approximation}

In the configuration model a node $i$ is characterized only by its degree $k_i$ so that its failure probability $\mathbb{P}(F\vert k_i)$ depends solely on this information. 
Hence the (average) final fraction of failed nodes is of the form
\begin{align}
 \rho^{*} = \sum_{k} p(k) \mathbb{P}(F\vert k).
\end{align}

The quantity $\rho^{*}$ allows for two different interpretations. 
On the system level, $\rho^{*}$ measures the final fraction of failed nodes, and $\mathbb{P}(F\vert k)$ the fraction of failed nodes with degree $k$. 
On the node level, $\rho^{*}$ can be seen as probability for a node to be failed, but if the node's degree $k$ is known, then its actual failure probability is given by $\mathbb{P}(F\vert k)$.
In the following, we proceed with successively decomposing $\mathbb{P}(F\vert k)$ into sums over products between conditional probabilities that assume more information about the network neighborhood, and probabilities that the neighborhood is in the assumed state. 

\begin{figure}[t]
 \centering
   \includegraphics[width=0.48\textwidth]{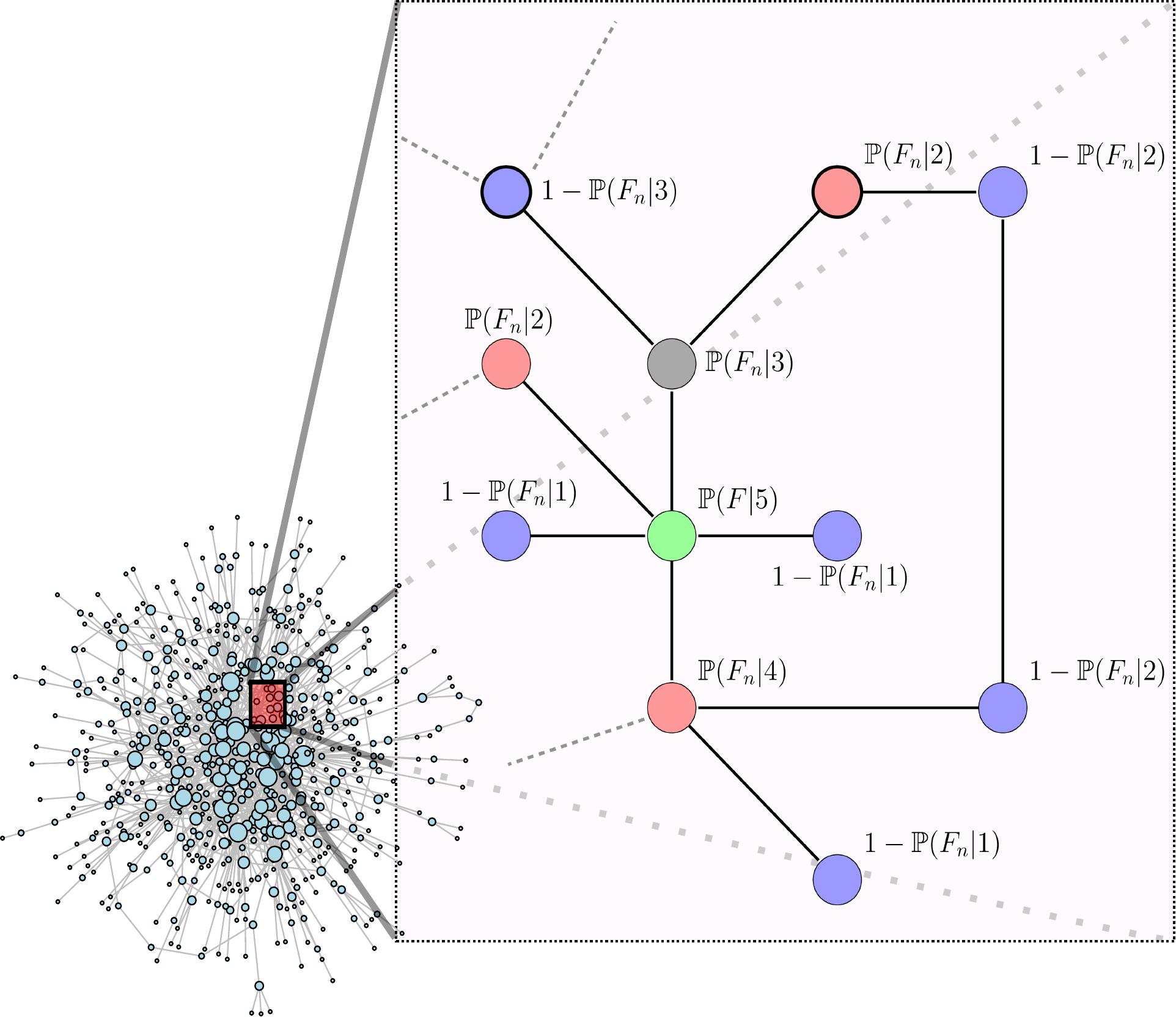}
 \caption{Illustration of the local tree approximation. 
 The green node is the focal node. 
 Its failure probability $\mathbb{P}(F\vert k=5)$ can be computed according to Equation~(\ref{eq:confailprob}), and depends on the state of its neighbors: Here, the two red ones and the gray one have failed, while the two blue ones are still functional. 
 The neighbors' conditional failure probabilities $\mathbb{P}(F_n\vert k)$ rely on the failure probabilities of their own neighbors without regarding the green focal point.}
\label{fig:illustration}
\end{figure}

\subsubsection{The conditional failure probability}

The computation of $\mathbb{P}(F\vert k)$ relies on the assumption of an infinite network size ($N \rightarrow \infty$), since the clustering coefficient for the configuration model vanishes in the limit, if the second moment of the degree distribution is finite~\cite{Newman2010}. 
Consequently, the topology simplifies to locally tree-like networks \cite{Dodds2009}, where neighbors of a node are not connected among each other, as illustrated in Fig.~\ref{fig:illustration}.
The node under consideration with degree $k$ is colored in green and is also called the focal node. 
Its failure probability $\mathbb{P}(F\vert k)$ decomposes into a sum over two factors, due to the combination of a locally tree-like network structure with the assumption that the local neighborhood defines the state of a node already:
\begin{align}\label{eq:confailprob}
 \mathbb{P}(F\vert k) = \sum^k_{n = 0} \mathbb{P}(F\vert k, n) b(n, k, \pi).
\end{align}
The factor $b(n, k, \pi)$ describes the general state of the neighborhood, namely the probability that among the $k$ neighbors of a node exactly $n$ have failed. 
The factor $\mathbb{P}(F\vert k, n)$ gives the probability that a node with degree $k$ fails after $n$ of its neighbors have failed. 
Therefore, it takes into account the ability of a node to withstand shocks (i.e., failing neighbors). 

If some of the neighbors of a node would be connected, which violates the local tree-like assumption, we would need to consider all possible (temporal) orders of their failures.
Instead, the configuration model allows for assigning every neighbor the same failure probability $\pi$, since no degree-degree correlations are present there, and each neighbor's failure is independent of the failure of the others because of the locally tree-like network structure.
Consequently, the number of failed neighbors of a node is binomially distributed so that $n$ neighbors can fail with probability
\begin{align*}
b(n, k, \pi) = \binom{k}{n} \pi^{n} (1-\pi)^{k-n}.
\end{align*}
However, the probability $\mathbb{P}(F\vert k, n)$ that such an event causes the failure of the considered node with degree $k$ may depend on specific properties of the neighbors, like e.g., their degrees $k_i$ and their failure probabilities $\mathbb{P}(F_n\vert k_i)$.
By allowing this dependence, we introduce a generalization of the existing heterogeneous mean field approximation which enables the analytical treatment of processes where failing nodes have different influences on their neighbors according to their degree. 

Also the conditional failure probability $\mathbb{P}(F\vert k, n)$ can be decomposed into the sum over two factors
\begin{align}
 \mathbb{P}(F\vert k, n) = \sum_{{\bf k_n} \in I^{n}} p({\bf k_n} \rvert k,n) \mathbb{P}(F\vert k,{\bf k_n}),
 \label{eq:condfprob}
\end{align}
where the sum runs over all possible configurations of neighbors' degrees denoted by $I$. $\bf k_n$ is an abbreviation for a vector $(k_1, \cdots, k_n)$ of failed neighbors' degrees and takes values in $I$.
Generally, $I = \mathbb{N}$ or $I = {[c]} := \{1,\cdots, c\}$ in the presence of a finite cutoff $c$.
Such a cutoff is inevitable in numerical computations or in the observation of (finite) real world systems, while it guarantees the finiteness of the second moment of $p(k)$.

The probability $p({\bf k_n} \rvert k,n)$ captures the precise state of the neighborhood given that exactly $n$ neighbors have failed. 
More precisely, it is the probability that the $n$ failed neighbors of a node with degree $k$ have degrees ${\bf k_n}$. 
This factor may depend on the failing probabilities of neighbors $\mathbb{P}(F_n\vert k_1), \cdots, \mathbb{P}(F_n\vert k_n)$, while conditioned on these failures, we denote $\mathbb{P}(F\vert k,{\bf k_n})$ the probability that a node with degree $k$ fails. 
The latter is determined by the specific cascading model, and will be discussed later for our two diversification models.

\subsubsection{The state of the neighborhood}

In order to compute the failure probability 
\begin{align}\label{eq:pfail}
 \mathbb{P}(F\vert k) = \sum^k_{n = 0} b(n, k, \pi)  \sum_{{\bf k_n} \in I^{n}} p({\bf k_n} \rvert k,n) \mathbb{P}(F \vert k,{\bf k_n})
\end{align}
we need to derive the state of the neighborhood as described by the average failure probability of a neighbor $\pi$, and the failure probability $p({\bf k_n} \rvert k,n)$ that the $n$ failed neighbors have degrees ${\bf k_n}$. 
Both depend on the degree distribution of a neighbor $p_n(k)$ as well as its failure probability $\mathbb{P}(F_n\vert k)$.

It is important to note that a neighbor with degree $k$ (illustrated by the gray node in Fig.~\ref{fig:illustration}) does not fail with probability $\mathbb{P}(F\vert k)$, since one of its links leads to the focal node (illustrated by the  green node in Fig.~\ref{fig:illustration}). 
Conditional on the event that the focal node has not failed yet, only the remaining ${k-1}$ neighbors of the gray neighbor (colored with bold fringe) could have caused the failure of the gray neighbor of the focal node. 
This model property is called the \emph{Without Regarding Property} (WOR) by Hurd and Gleeson~\cite{Hurd2013}. 
Therefore, a neighbor's failure probability $\mathbb{P}(F_n\vert k)$ is
\begin{align}\label{eq:pfailnb}
\mathbb{P}(F_n\vert k) =  \sum^{{k-1}}_{n = 0} b(n, {k-1}, \pi) \sum_{{\bf k_n} \in I^{n}}  p({\bf k_n} \rvert k,n) \mathbb{P}(F\vert k,{\bf k_n}).
\end{align}

It remains to calculate $p({\bf k_n} \rvert k,n)$ and the unconditional failure probability of a neighbor $\pi$. 
Both depend on the degree distribution $p_n(k)$ of a neighbor. 
Because of the local tree approximation, it is independent of the degree distribution of the other nodes in the network:
\begin{align}
 p_n(k) := \frac{k p(k)}{z},
\end{align}
where $z := \sum_{k} k p(k)$ denotes the normalizing average degree.
$p_n(k)$ is proportional to the degree $k$ in the configuration model, because each of a neighbor's $k$ links could possibly connect the neighbor with the focal node (see, e.g., \cite{Newman2010}).

We, therefore, obtain the unconditional failure probability $\pi$ of a neighbor by
\begin{align}\label{eq:pi}
 \pi = \sum_{k} p_n(k) \mathbb{P}(F_n\vert k) = \sum_{k} \frac{k p(k)}{z} \mathbb{P}(F_n\vert k).
\end{align}
Similarly, the degree distribution of a neighbor conditional on its failure can be written as $\mathbb{P}(F_n\vert k) p(k) k /z \pi$ so that we can calculate the probability $p({\bf k_n} \rvert k,n)$ that the $n$ failed neighbors have degrees $\bf k_n$ by
\begin{align}\label{eq:nbdegrees}
p({\bf k_n} \rvert k,n) = \prod^{n}_{j = 1} \frac{p(k_j) k_j \mathbb{P}(F_n\vert k_j)}{z \pi},
\end{align}
since the neighbors are independent of each other, according to the locally tree-like network structure.

\subsubsection{Fixed point iteration for the conditional failure probability}
In short, the vector ${\bf\mathbb{P}(F_n\vert k)} = \left( \mathbb{P}(F_n\vert k)\right)_{k \in \{1,\cdots, c\}}$ turns out to be a fixed point of a vector valued function ${\bf L\left( p \right)}$ so that for the $k-$th component we have:
\begin{align*}\label{eq:fixp}
\begin{split}
 \mathbb{P}(F_n\vert k)  = & \sum^{{k-1}}_{n = 0} b(n, {k-1}, \pi)  \sum_{{\bf k_n} \in I^{n}} \mathbb{P}(F\vert k,{\bf k_n}) \cdot \\
 & \cdot \prod^{n}_{j = 1} \frac{p(k_j) k_j \mathbb{P}(F_n\vert k_j)}{z \pi} 
  =  L_{k}\left( {\bf\mathbb{P}(F_n\vert k)} \right).
\end{split} 
\end{align*}
Such a fixed point exists according to the Knaster-Tarski Theorem, since ${\bf L\left( p \right)}$ is monotone with respect to a partial ordering and maps the complete lattice ${[0,1]}^{I}$ onto itself.~\footnote{Definition of the partial ordering: Two vectors  $\bf x$,  $\bf y$ $\in {[0,1]}^{I}$ are ordered as $\bf x \leq y$, if and only if $x_i {\bf \leq} y_i$ holds for all their components $i \in I$.}

Thus, starting from an initial vector ${\bf \mathbb{P}(F_n\vert k)^{(0)}}$ which is defined by the considered cascading model, we can compute the fixed point iteratively by 
\begin{align}
 {\bf \mathbb{P}(F_n\vert k)^{(t+1)}} = {\bf L\left( {\bf \mathbb{P}(F_n\vert k)^{(t)}} \right)},
\end{align}
with
\begin{align}
\pi^{(t)} = \sum_{k} p_n(k) \mathbb{P}(F_n\vert k)^{(t)}.
\end{align}
Each iteration step $(t)$ corresponds to one discrete time step of the cascading process so that 
\begin{align}
 \rho^{(t)} = \sum_k p(k) \mathbb{P}(F\vert k)^{(t)}
\end{align}
can be interpreted as average fraction of failed nodes in the network at time $t$.
Note that the relation between $\bf \mathbb{P}(F\vert k)^{(t)}$ and $\bf \mathbb{P}(F_n\vert k)^{(t)}$ is described by Equation (\ref{eq:pfail}) and Equation (\ref{eq:nbdegrees}).

\subsubsection{Simplification for homogeneous failure probability}\label{sec:anaInward}

In case the impact of a failing neighbor does not depend on its degree, the failure probability $\mathbb{P}(F\vert k,{\bf k_n}) = \mathbb{P}(F\vert k,n)$ does not depend on the degrees ${\bf k_n}$ of its $n$ failed neighbors and Equation~(\ref{eq:fixp}) can be simplified to
\begin{align}
 \begin{split}
 \mathbb{P}(F_n\vert k) = & \sum^{{k-1}}_{n = 0}  b(n, {k-1}, \pi) \mathbb{P}(F\vert k,n) \\
 & \prod^{n}_{j = 1} \frac{1}{\pi} \sum_{k_j \in I} \frac{p(k_j) k_j \mathbb{P}(F_n\vert k_j)}{z} \\ 
 = & \sum^{{k-1}}_{n = 0} b(n, {k-1}, \pi) \mathbb{P}(F\vert k,n),
 \end{split}
\end{align}
using Equation (\ref{eq:pi}).
Inserting this into Equation~(\ref{eq:pi}) leads to the fixed point equation 
\begin{align}\label{eq:scalarfixp}
 \pi = \sum_{k} \frac{k p(k)}{z} \sum^{{k-1}}_{n = 0} b(n, {k-1}, \pi) \mathbb{P}(F\vert k,n)
\end{align}
which in this case involves the scalar $\pi$ instead of the vector ${\bf\mathbb{P}(F_n\vert k)}$ in Equation~(\ref{eq:fixp}). 
With this information the final fraction of failed nodes can be computed as
\begin{align}
 \rho^{*} = \sum_{k} p(k) \sum^k_{n = 0} b(n, k, \pi) \mathbb{P}(F\vert k,n),
\end{align}
as already known from the literature \cite{Gleeson2007, Dodds2009}.
Still, this simpler approach is not able to capture the cascade dynamics of the damage diversification model.

\subsubsection{The ability of a node to withstand a shock}The only piece missing in our derivations is the model specific probability $\mathbb{P}(F\vert k,{\bf k_n})$ that a node with degree $k$ fails exactly after $n$ of its neighbors with degrees ${\bf k_n}$ have failed. 
This probability captures the failure dynamics, and is thus defined by a node's fragility $\phi(k, {\bf k_n})$ and its threshold $\Theta(k)$. 
The given information about the degrees $k, {\bf k_n}$ can in principle enter both variables, although the cumulative threshold distribution $F_{\Theta(k)}$ tends to depend solely on properties of the node itself, e.g. the degree $k$. 
Because a node fails, if its fragility exceeds its threshold, we have
\begin{align}
 \mathbb{P}(F\vert k,{\bf k_n}) = \mathbb{P} \left(\Theta(k) \leq \phi(k, {\bf k_n})\right). \label{eq:18}
\end{align}

More generally, with respect to known weight distributions $p_{W(k_j,k)}$ for given degree $k_j$ of a neighbor and the degree $k$ of a node, Equation~(\ref{eq:18}) reads:
\begin{align}\label{eq:abilShock}
\begin{split}
 & \mathbb{P}(F\vert k,{\bf k_n})  = \mathbb{P} \left(\Theta(k) \leq \sum^n_{j=1} W(k_j,k)\right) \\ & = \int F_{\Theta(k)}(w) \left( p_{W(k_1,k)} * \cdots * p_{W(k_n,k)}\right)(w) \mathop{dw}.
\end{split}
\end{align}
The last equation holds if the weight distributions $p_{W(k_j,k)}$ are independent, and the $*$ denotes a convolution. 

In the simpler case of our two model variants, the weights $W(k_j,k)$ are completely deterministic. 
In accordance with the definition of the weights in Equation~(\ref{eq:weights}) we calculate for the ED case
\begin{align}
 \mathbb{P}(F\vert k,{\bf k_n})^{\rm{(ED)}} = \mathbb{P}(F\vert k, n)^{\rm{(ED)}} = F_{\Theta (k)}\left(\frac{n}{k}\right) 
\end{align}
which is independent of the neighbors' degrees. Consequently, the calculation of the average final fraction of failed nodes can be simplified as outlined in Section~\ref{sec:anaInward}.

For the DD case the fixed point iteration needs to take into account all degrees of the failed neighbors, since they define the loss $1/k_j$ the focal node faces. 
Thus, we have
\begin{align}
\mathbb{P}(F\vert k,{\bf k_n})^{\rm{(DD)}} =  F_{\Theta(k)}\left(\sum^{n}_{j=1} \frac{1}{k_j}\right).
 \end{align}

\subsubsection{DD case: Correct Heterogeneous Mean Field Approximation (cHMF)}

\begin{figure}[t]
 \centering
\includegraphics[width=0.24\textwidth]{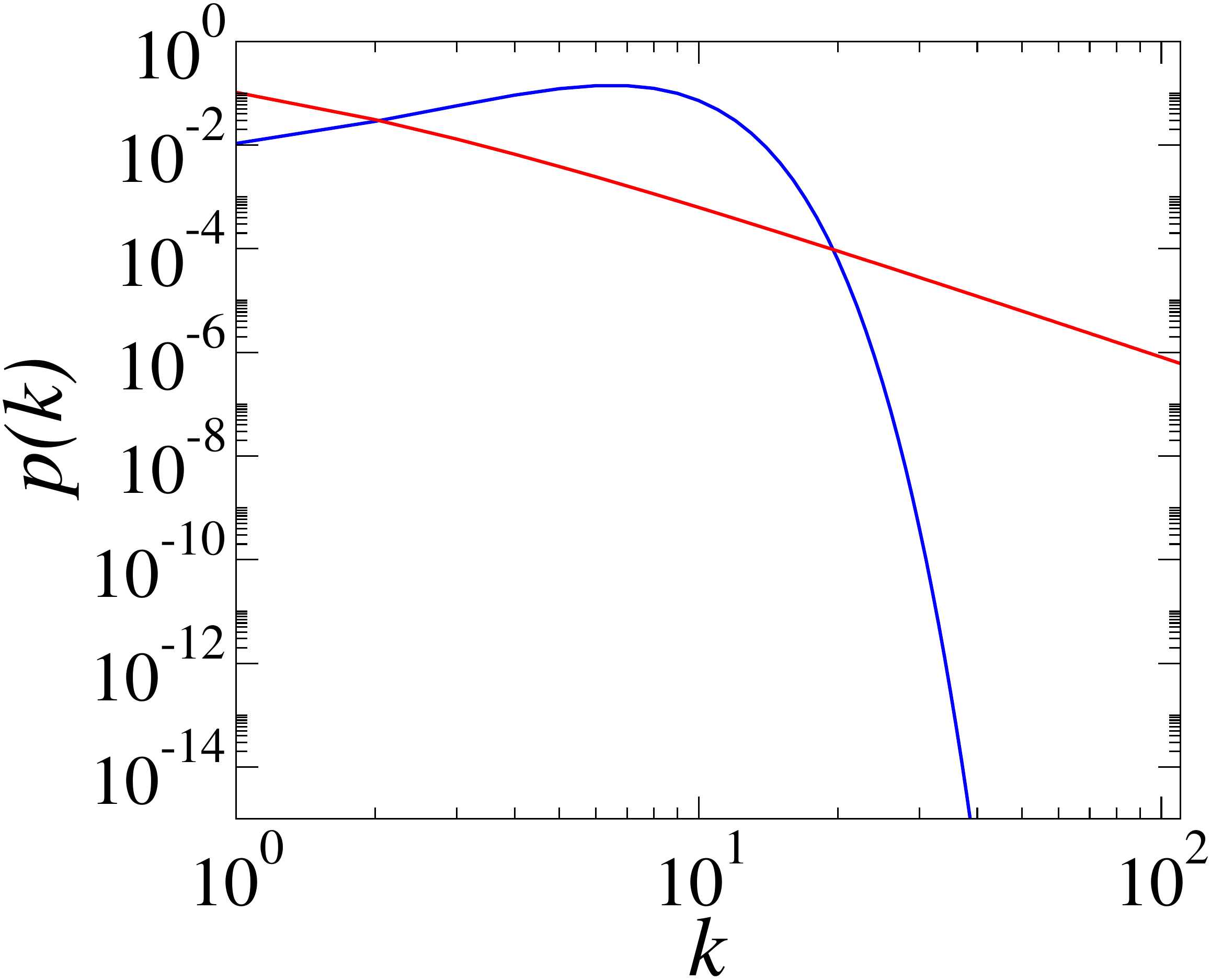}
\includegraphics[width=0.23\textwidth]{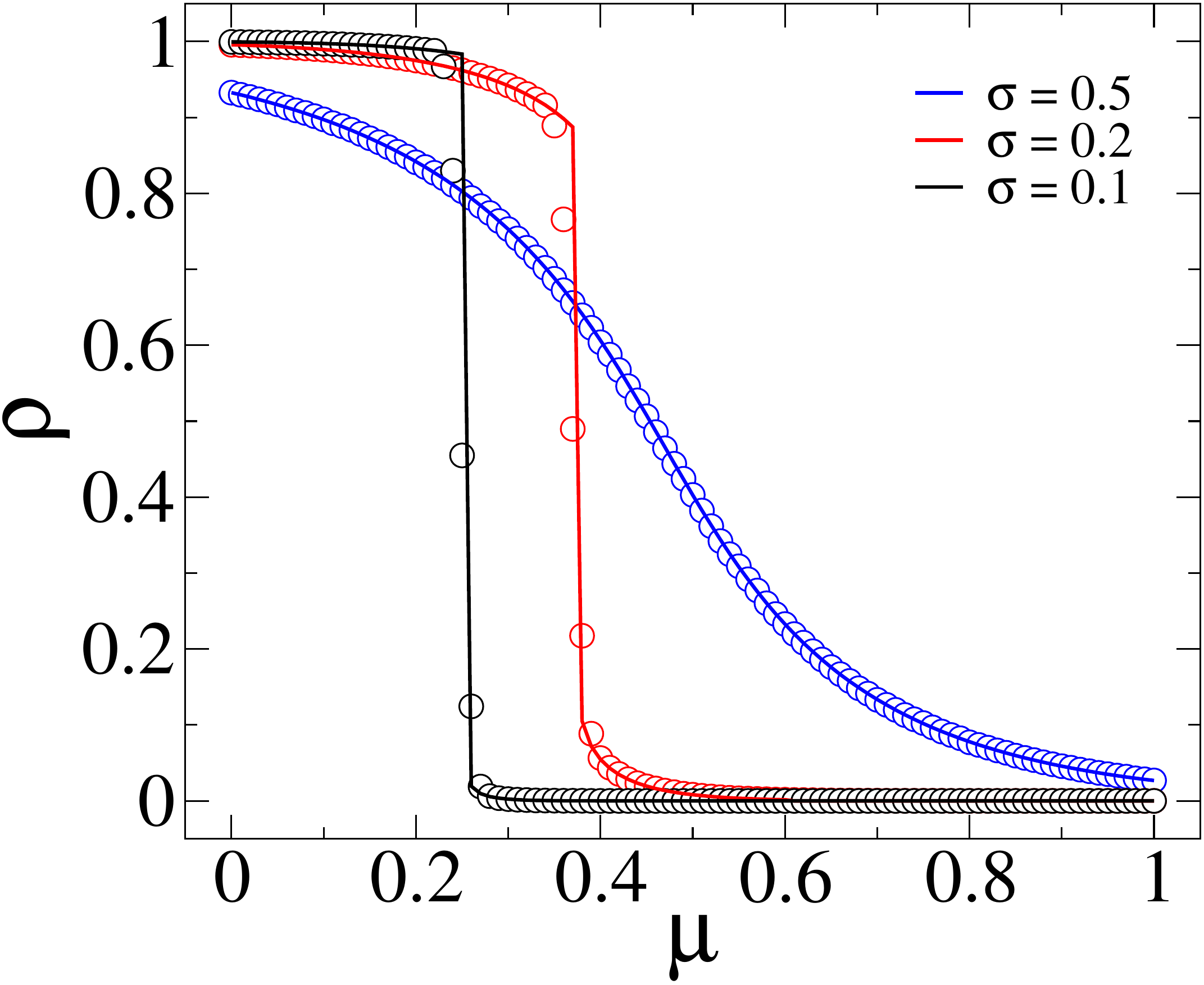}
\begin{picture}(0,0)
    \put(-248,90){(a)}
    \put(-122,90){(b)}
   \end{picture}
 \caption{(a) The studied degree distributions: Poisson distribution with parameter ${\lambda=8}$ and cutoff degree ${c = 50}$ (blue), scale free distribution with exponent ${\gamma = 3}$ and maximal degree ${c = 200}$ (red) in log-log scale. (b) Comparison of the final fraction of failed nodes obtained by simulations (symbols) on networks with $1000$ nodes as average over $2000$ independent realizations with numerical results from the cHMF (lines) for the DD case on Poisson random graphs. The thresholds $\Theta$ are normally distributed with mean $\mu$ and standard deviation $\sigma$ ($\Theta \sim \mathcal{N}(\mu, \sigma)$).}
\label{fig:degree}
\end{figure}
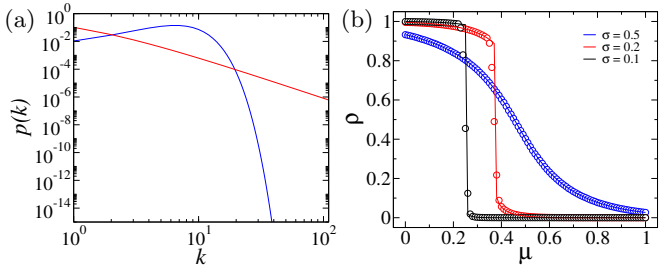
The probability of the failure of a node or neighbor with degree $k$ and $n$ failed neighbors $\mathbb{P}(F\vert k,n)^{\rm{(DD)}}$ given by Equation~(\ref{eq:condfprob})
needs to be re-calculated for each fixed point iteration step.
For the DD case, this involves the calculation of the convolution of the impact distribution $p_{\rm{imp}}$ of a failed neighbor, which depends on the iteratively updated failure probability $\mathbb{P}(F_n\vert k)$.
More precisely, one failed neighbor inflicts the loss $1/k$ with probability \begin{align}
 p_{\rm{imp}}\left(\frac{1}{k}\right) = \mathbb{P}(F_n\vert k)\frac{k p(k)}{z \pi}
\end{align}
(that is conditioned on its failure) independently of the other failed neighbors.
Thus, the fragility $\phi(k,n)$ of a node with degree $k$ and $n$ failed neighbors is distributed according to the $n$-th convolution of this impact distribution $p^{*n}_{\rm{imp}}$ and we have
\begin{align*}
\mathbb{P}(F\vert k, n) & = \mathbb{P} \left(\Theta(k) \leq \phi(k, n)\right)  =  \sum_{l}p^{*n}_{\rm{imp}}(l) F_{\Theta}(l),
\end{align*}
where the inner sum runs over all possible values of the fragility.

Since the convolutions are computationally demanding (in terms of time and especially memory), we approximate $p^{*n}_{\rm{imp}}$ by first binning it to an equidistant grid and then using Fast Fourier Transformations (FFT) in order to take advantage of the fact that convolutions correspond to simple multiplications in Fourier space (see for instance \cite{Ruckdeschel2010,Frigo2005}). 
Although this is numerically accurate enough for the calculation of the final fraction of failed nodes, for the reporting of the vectors $\bf \mathbb{P}(F\vert k)$ and  $\bf \mathbb{P}(F_n\vert k)$ we use a more precise direct convolution of the binned impact distributions $p_{\rm{imp}}$, as is described in the Appendix. 
Fig. \ref{fig:degree} shows that our numerical results coincide with simulations.

\subsubsection{Neglecting the neighbors' degrees in the failure probability (simpHMF)}

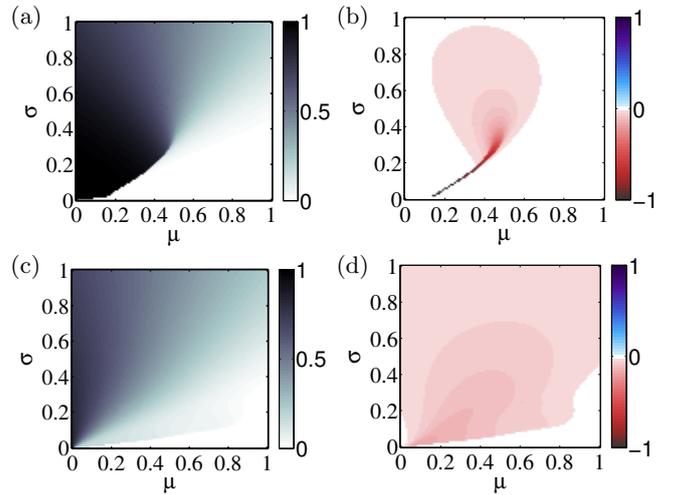
\begin{figure}[t]
 \centering
\includegraphics[width=42mm]{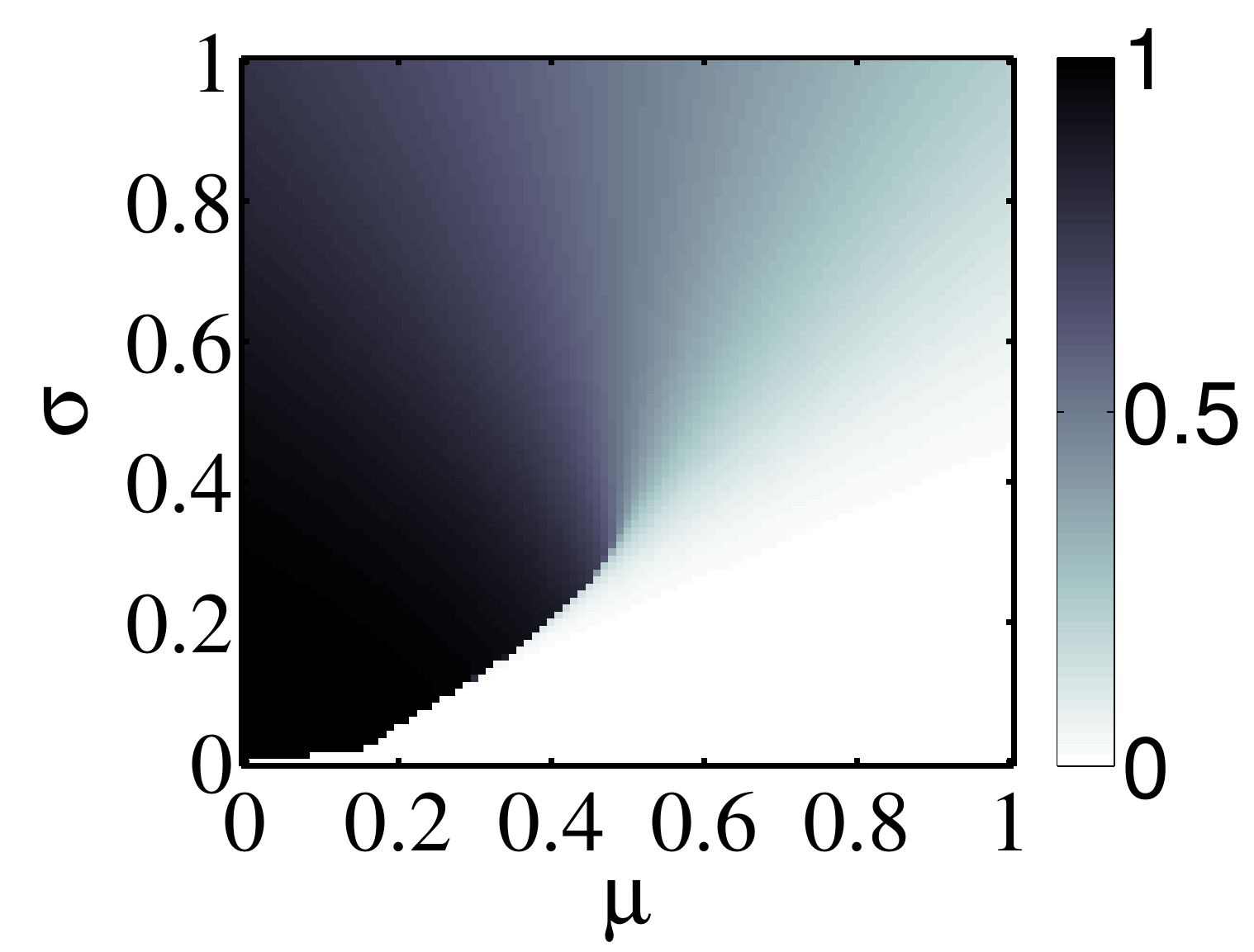}
\includegraphics[width=42mm]{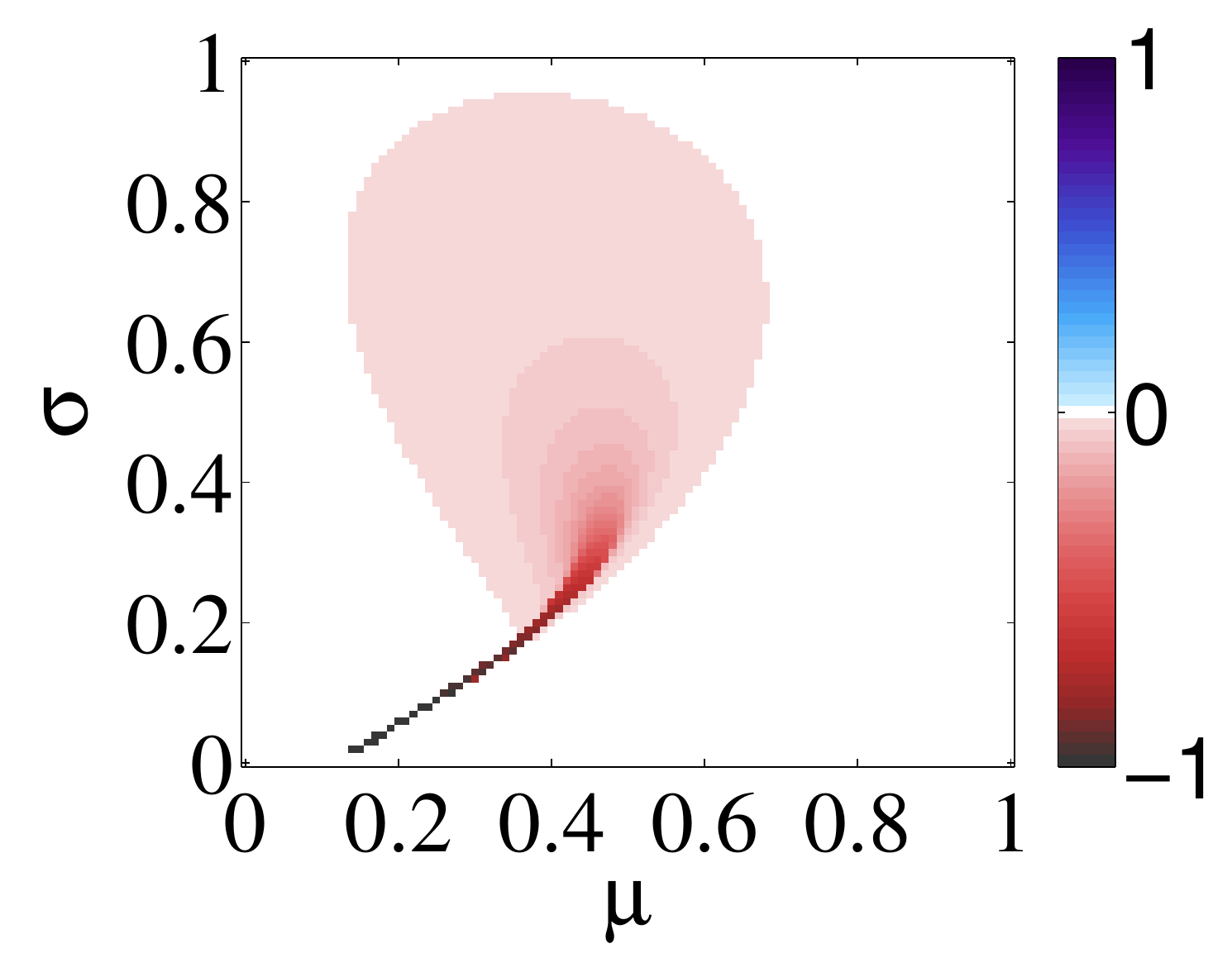}
\includegraphics[width=42mm]{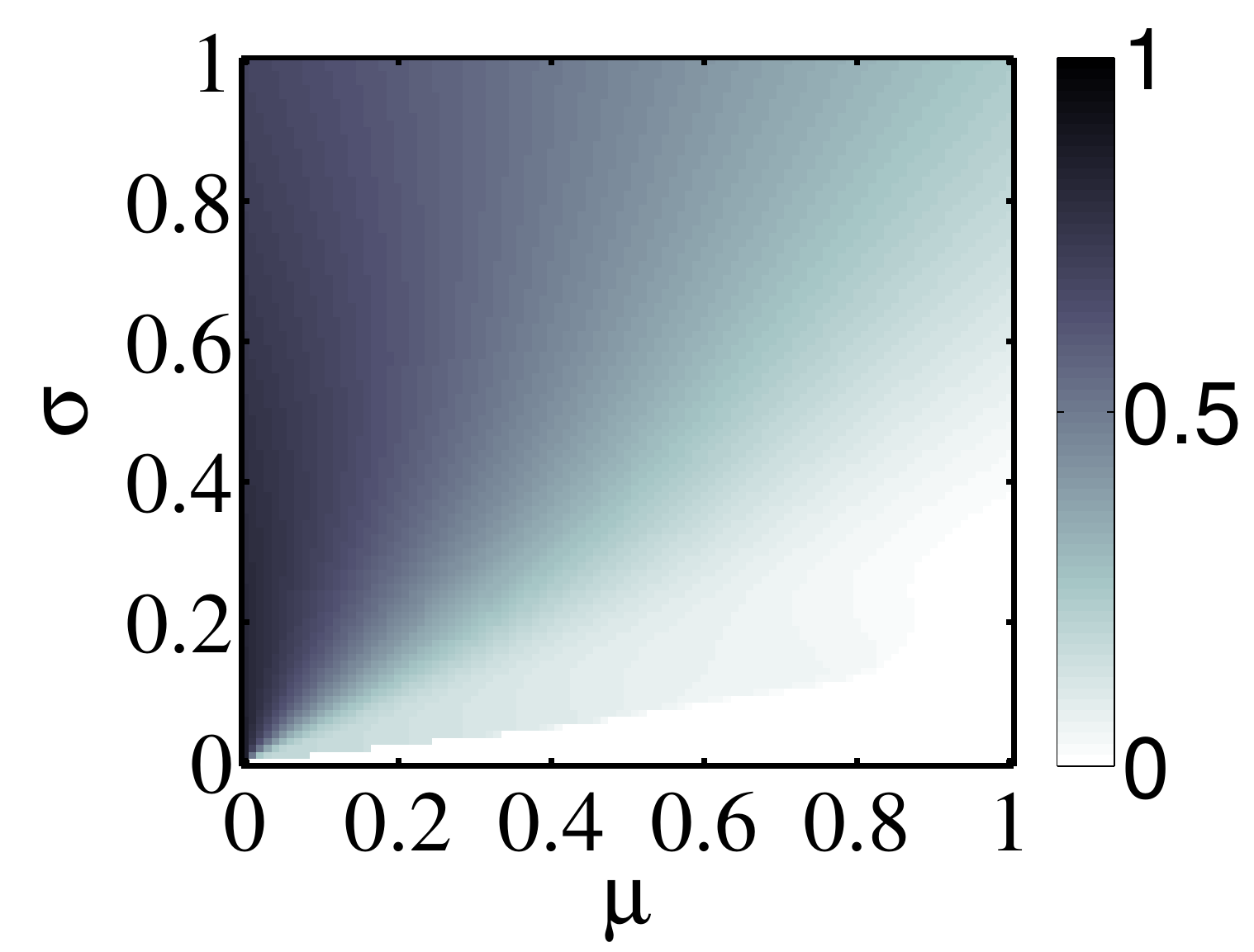}
\includegraphics[width=42mm]{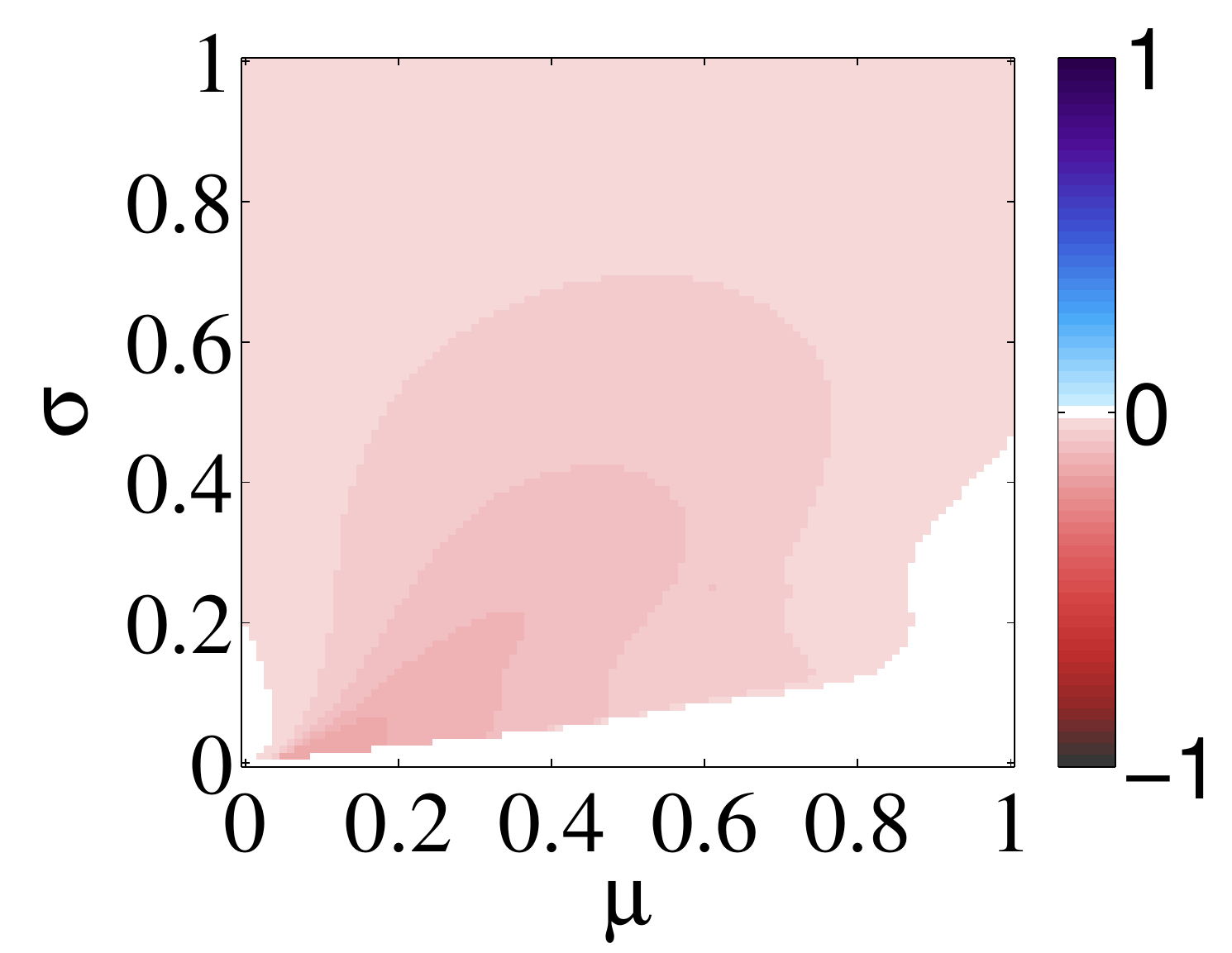}
\begin{picture}(0,0)
    \put(-245,178){(a)}
    \put(-123,178){(b)}
    \put(-245,84){(c)}
    \put(-123,84){(d)}
   \end{picture}

 \caption{(a) Fraction of failed nodes obtained by (simpHMF) for the DD case on Poisson random networks with $\lambda = 8$ and $c = 50$. (b) Difference between the correct version (cHMF) and a). (c) Fraction of failed nodes obtained by (simpHMF) for the DD case on scale free networks with ${\gamma = 3}$ and $c = 200$. (d) Difference between the corresponding correct version (cHMF) version and c). The thresholds $\Theta$ are normally distributed with mean $\mu$ and standard deviation $\sigma$ ($\Theta \sim \mathcal{N}(\mu, \sigma)$).}
 \label{fig:simpHMF}
\end{figure}

Considering the computational complexity of the cHMF approach we described above, it is worth asking whether we can approximate it with a simpler version as, e.g., the one described in Section \ref{sec:anaInward}, and still obtain reasonably good results. 

This would require the probability $\mathbb{P}(F\vert k,{\bf k_n})$ to be independent of the neighbors' degree. Consequently, we would assume that every failed neighbor transfers a load $1/k$ with probability \begin{align}
 p_{\rm{simp}}\left(\frac{1}{k}\right) = \frac{k p(k)}{z},
\end{align}
although its degree does not need to coincide with $k$.
Therefore, by computing
\begin{align}
\mathbb{P}(F\vert k,{\bf k_n})= \sum_{l}p^{*n}_{\rm{simp}}(l) F_{\Theta}(l)  = \mathbb{P}(F\vert k, n) 
\end{align}
we can calculate the failure probability
initially without the need to update it in each fixed point iteration.
Although this approach is more convenient, as shown in Fig.~\ref{fig:simpHMF}, it is inadequate for the damage diversification variant, especially in combination with skew degree distributions (as, e.g., in case of scale free networks). 
This is because if we follow this simplified calculation, we lose the risk reducing effect by hubs that are connected to more nodes, and we would draw opposite conclusions about systemic risk. 
So, as shown in Fig.~\ref{fig:simpHMF}, it is crucial to use the correct HMF to explain our simulation results.

\section{Numerical Results}\label{sec:results}
We calculate the final fraction of failed nodes for Poisson random graphs and scale free networks with degree distributions
\begin{align}\label{eq:degrdistr}
 p_P(k) := \frac{1}{S_P} \frac{\lambda^k}{k!}, \ \ \   p_S(k) := \frac{1}{S_S} \frac{1}{k^{\gamma}}
\end{align}
for $k \in \{1, \cdots, c\}$ with normalizing constants \begin{align}
 S_P := \sum^{c}_{k=1} \frac{\lambda^k}{k!} \ \ \ \text{and} \ \ \ S_S := \sum^{c}_{k=1} \frac{1}{k^{\gamma}},
\end{align}
as shown in Fig.~\ref{fig:degree}(a).

The Poisson random graphs are of interest as limit of the well studied Erd\"os-R\'enyi random graphs \cite{Erdos1959}, and  in comparison to the simulations serve as benchmark for our method (see Fig.~\ref{fig:degree}(b)), while the class of scale free networks is more realistic with respect to real world networks~\cite{Nationalbank2003}.

Similar to \cite{Watts2002SimpleModelof} and \cite{Gleeson2007} we study normally distributed thresholds $\Theta \sim \mathcal{N}(\mu,\sigma^2)$ with mean $\mu$ and standard deviation $\sigma$, but, we explore the role of the thresholds' heterogeneity as well as their mean size more extensively.
Although our analytic framework also applies to more general cases, here we assume the thresholds to be independent from the degree $k$ of a node.

\begin{figure*}[htb]
\centering
  \begin{tabular}{@{}cccc@{}}
 & \bf ED & & \bf DD\\
 \bf Poisson &
\includegraphics[valign=m,width=.3\textwidth]{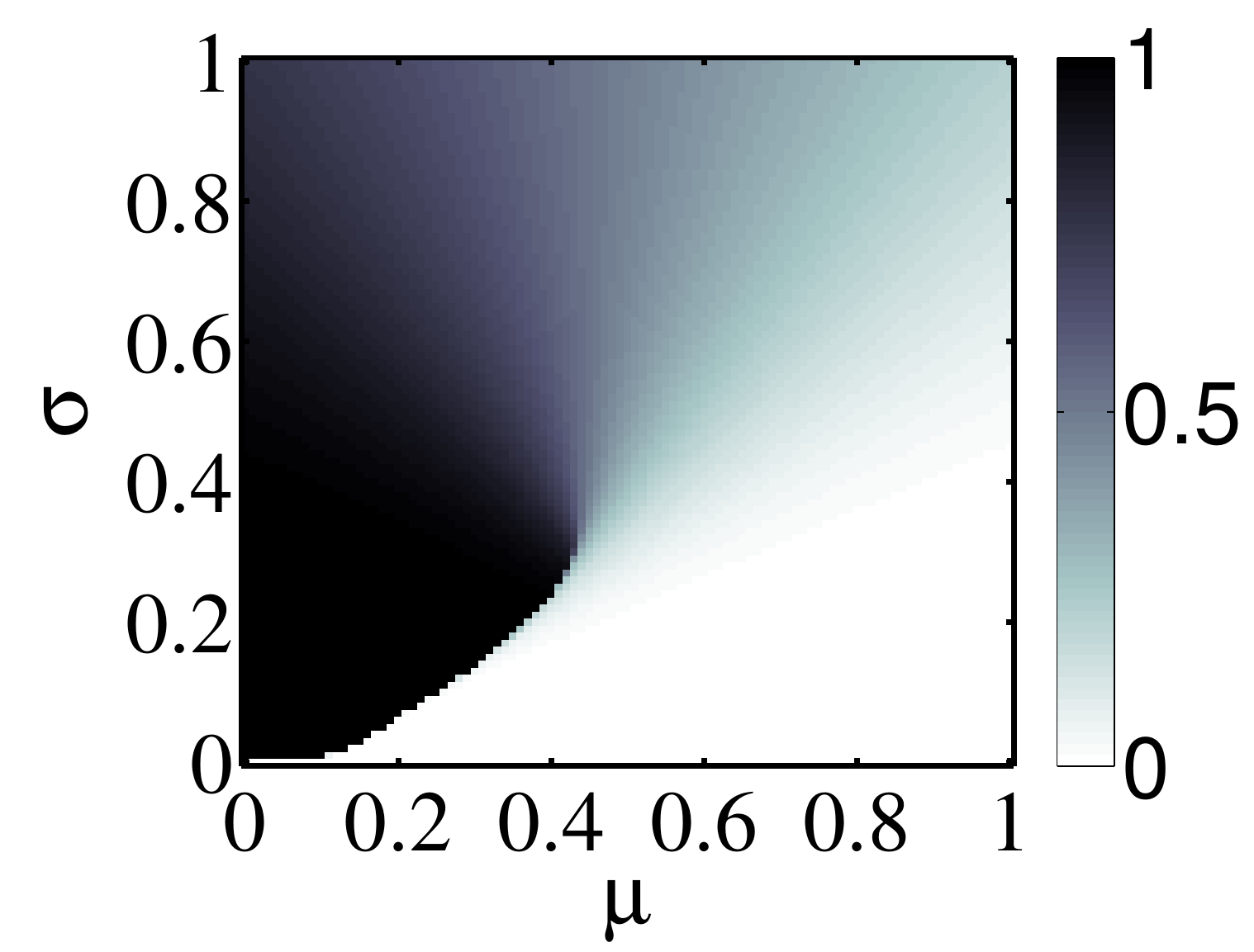} &
    \includegraphics[valign=m,width=.3\textwidth]{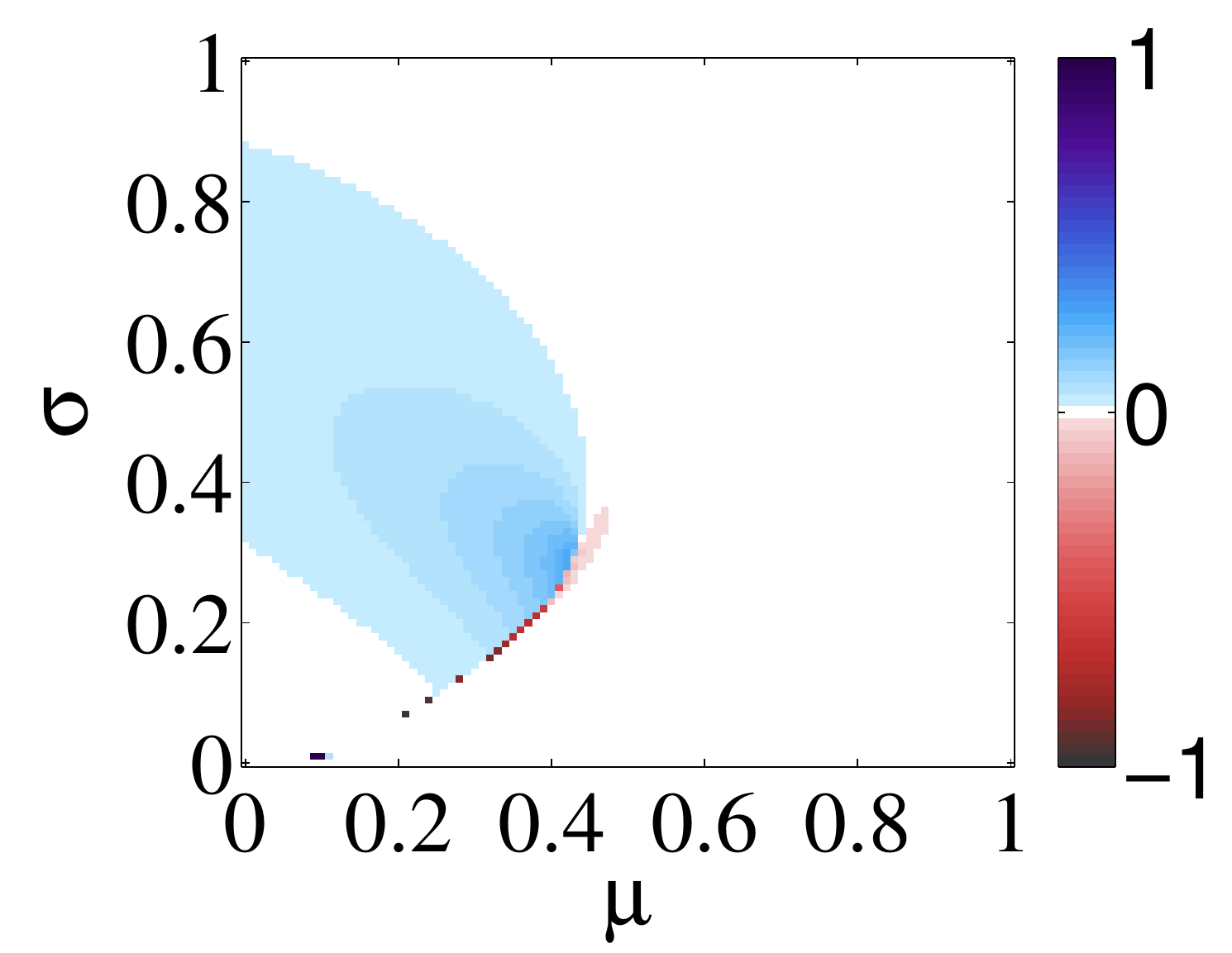} &
    \includegraphics[valign=m,width=.3\textwidth]{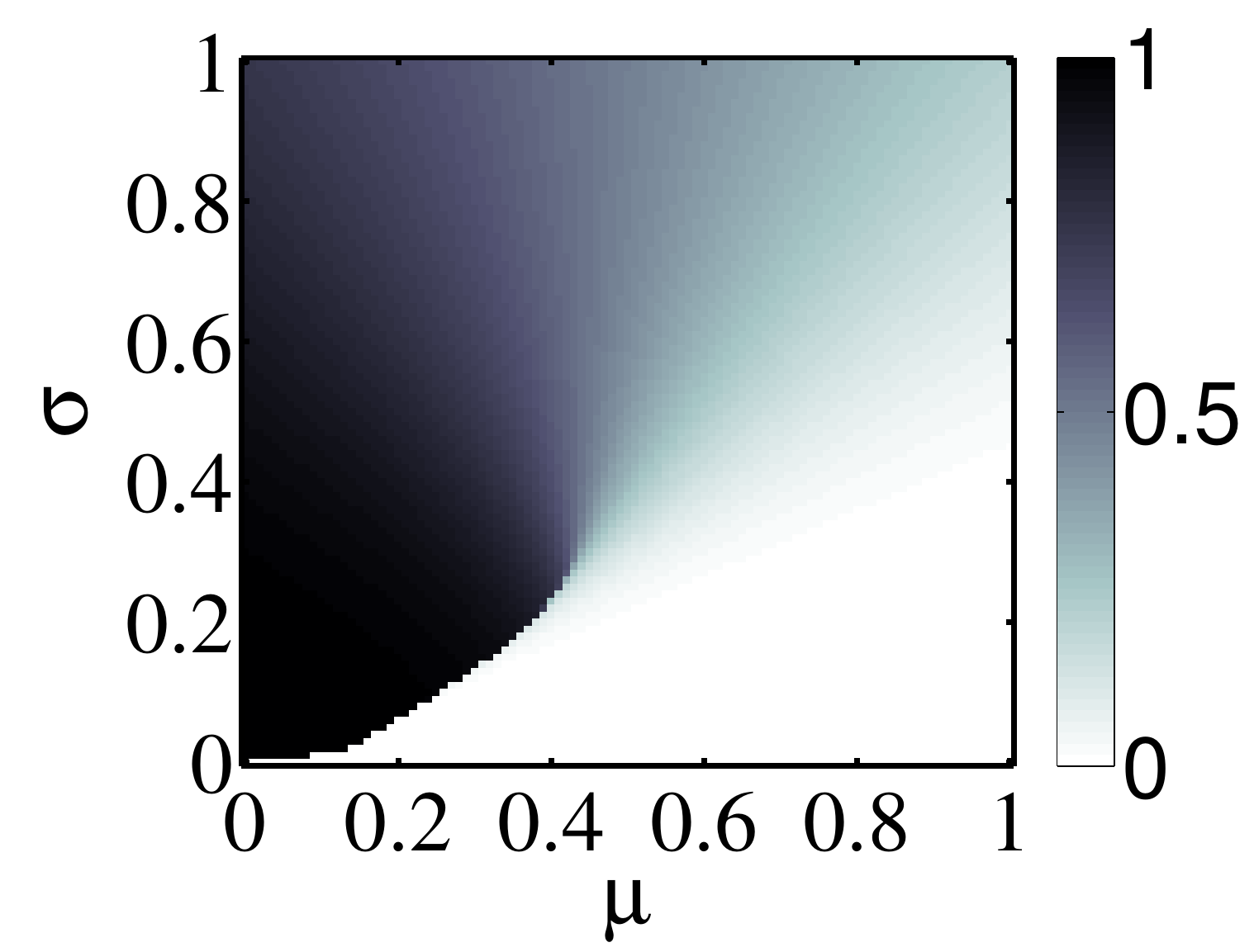}   \\

  &
    \includegraphics[valign=m,width=.3\textwidth]{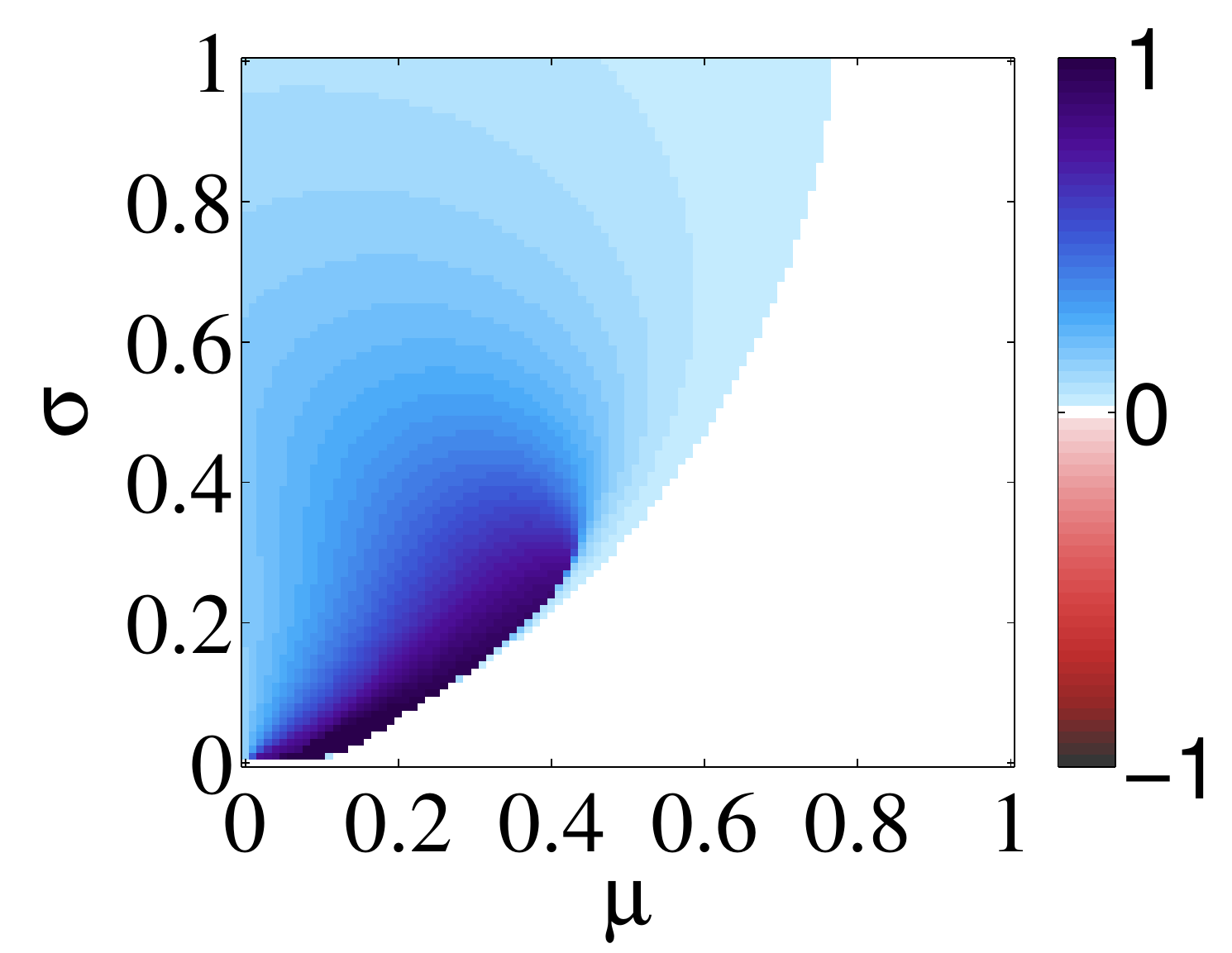} &
    \includegraphics[valign=m,width=.3\textwidth]{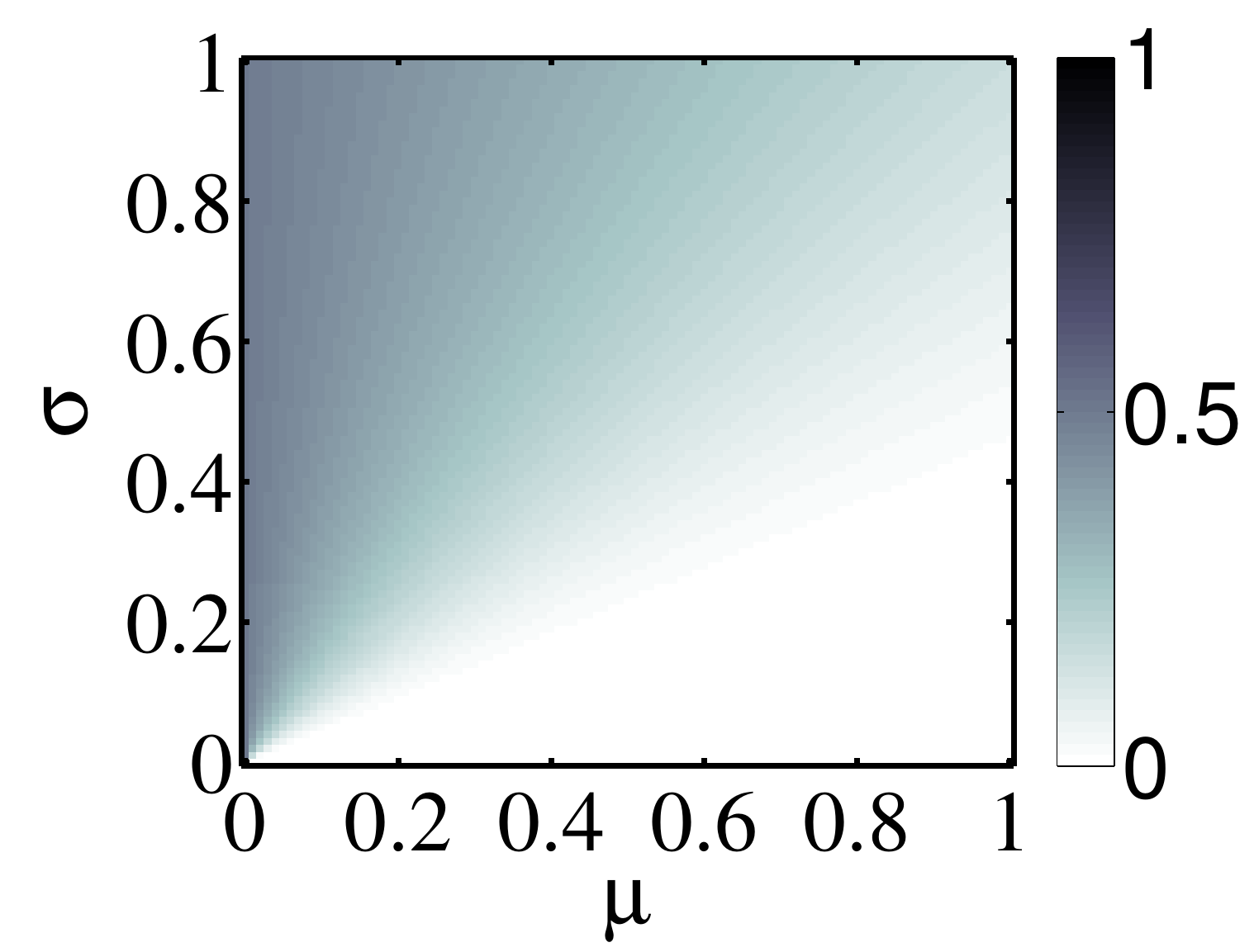} &
    \includegraphics[valign=m,width=.3\textwidth]{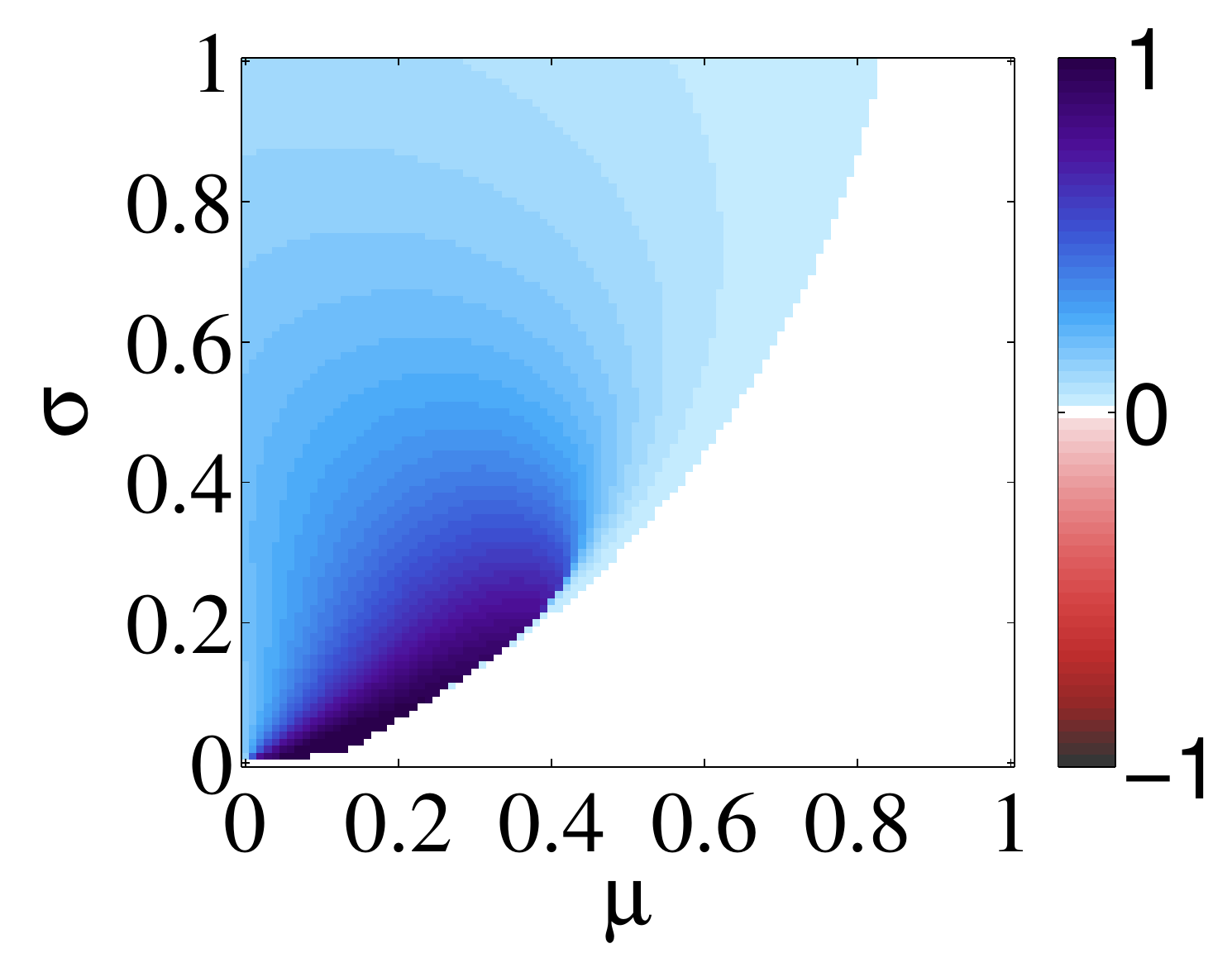}   \\
   \bf Scale free &
    \includegraphics[valign=m,width=.3\textwidth]{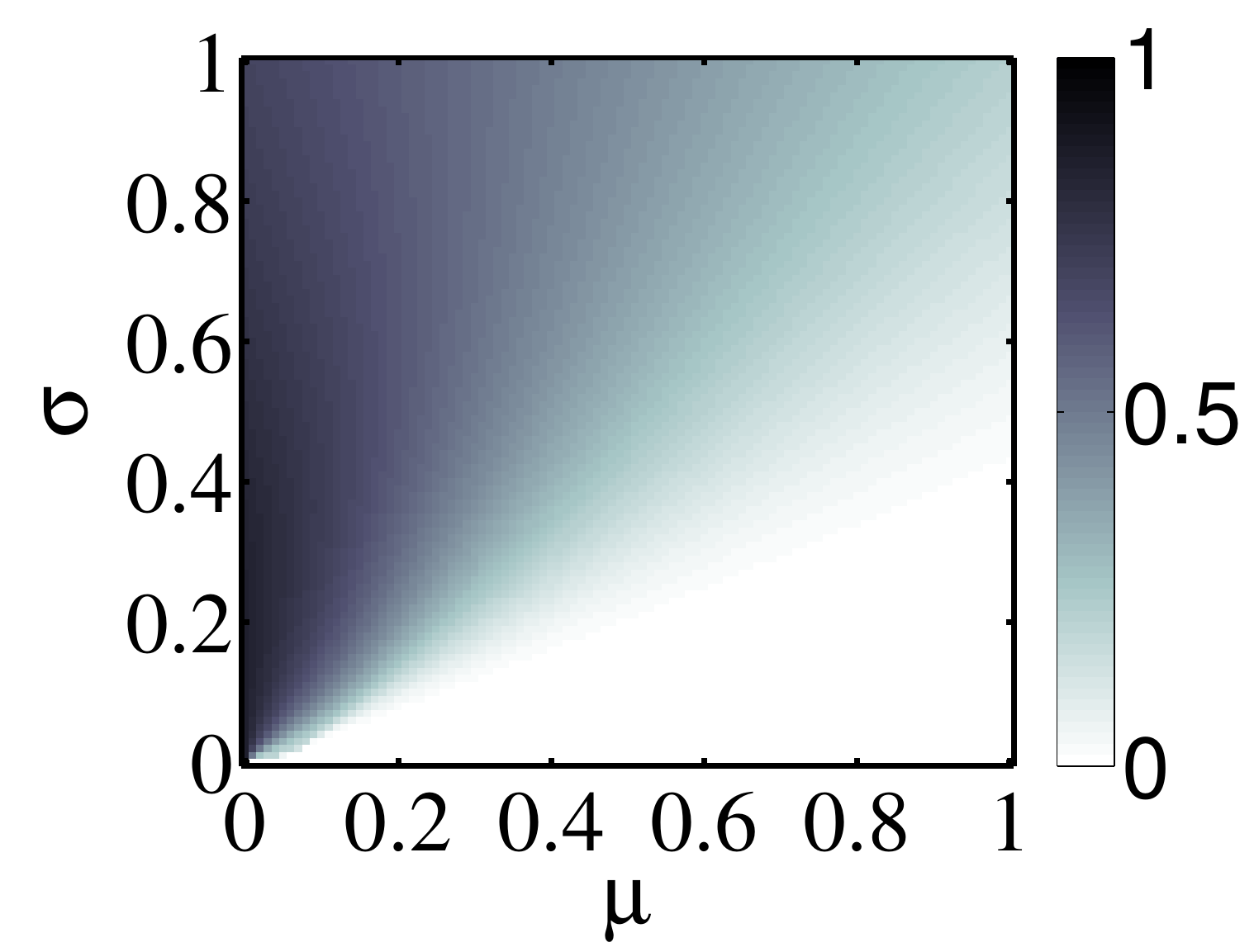} &
    \includegraphics[valign=m,width=.3\textwidth]{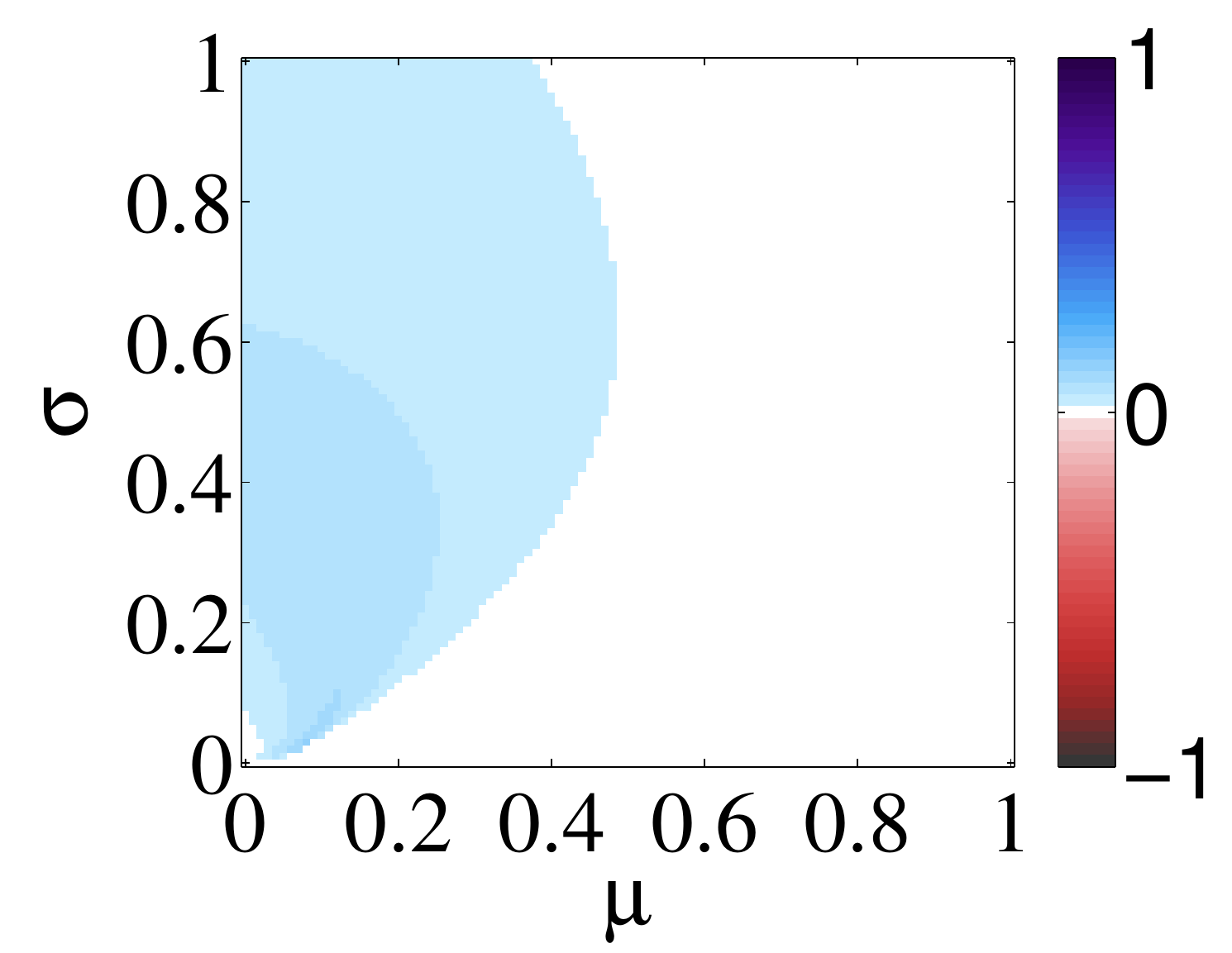} &
    \includegraphics[valign=m,width=.3\textwidth]{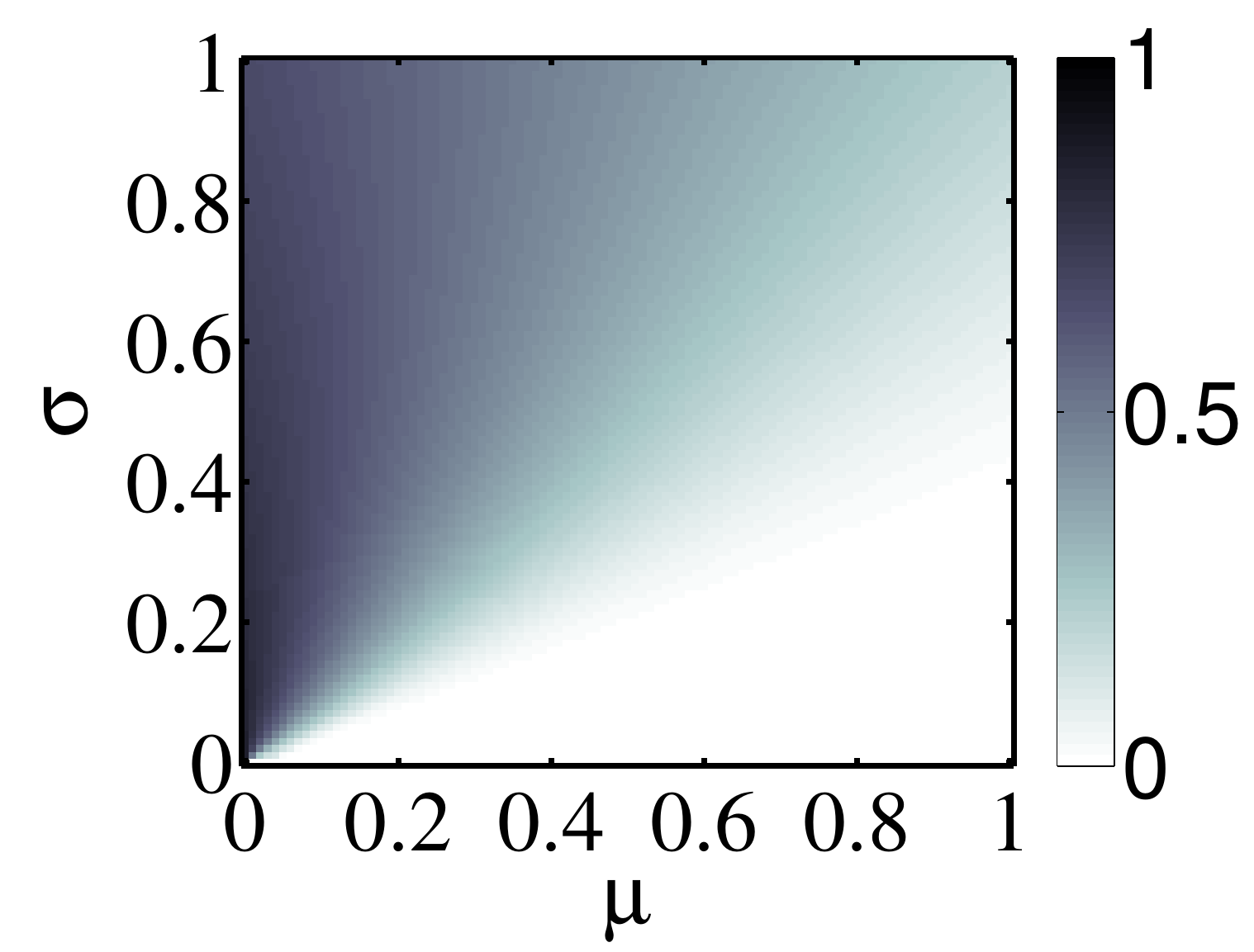}  \end{tabular}
  \caption{Phase diagrams for the final fraction of failed nodes $\rho$ for different degree distributions, diversification variants, and their differences. 
  The thresholds $\Theta$ are normally distributed with mean $\mu$ and standard deviation $\sigma$ ($\Theta \sim \mathcal{N}(\mu, \sigma)$). 
  The darker the color the higher is the systemic risk. 
  First row: Poisson distribution with parameter ${\lambda=8}$ and cutoff degree ${c = 50}$ for the ED (left) and DD (right). 
  The middle shows their difference ${\rho^{\rm{(ED)}}-\rho^{\rm{(DD)}}}$.
  Third row: Scale free distribution with exponent ${\gamma = 3}$ and maximal degree ${c = 200}$ for the ED (left) and DD (right).
  The middle shows their difference ${\rho^{\rm{(ED)}}-\rho^{\rm{(DD)}}}$.
  Second row: The difference between the diagrams with Poisson and Scale free degree distributions for the ED variant (left).
  Similarly for the DD variant (right).
  In the middle the initial fraction of failed nodes $\rho(0)$ is illustrated. ${\rho_o := \rho(0)}$ is constant along the lines ${\sigma = \mu / \Phi^{-1}(\rho_0)}$.}
  \label{fig:summary}
\end{figure*}

\subsection{System Failures} 
Complementary to \cite{Watts2002SimpleModelof} we find that increasing the standard deviation can also reduce the cascade size - even though more nodes fail initially, similarly to what was reported for fully connected networks in \cite{Lorenz2009b}, where the ED and DD cases coincide. 
The initial fraction of failed nodes is determined by the nodes with negative thresholds ($\Theta \leq 0$) and is thus given by $F_{\Theta} (0) = \Phi(\frac{-\mu}{\sigma})$, where $\Phi$ denotes the cumulative distribution function of the standard normal distribution, as illustrated in the center of Fig.~\ref{fig:summary}. 

More pronounced is the sudden regime shift from a region of negligible systemic risk to an almost complete break-down of the system that occurs for Poisson random graphs for moderate robustness ($\mu < 0.4$) and increasing threshold heterogeneity (measured by $\sigma$) as shown in the first row of Fig.~\ref{fig:summary}. 
The same phenomenon can be found in fully-connected or regular random networks \cite{Lorenz2009b}, but for a more realistic degree distribution, like the scale free distribution with exponent $\gamma = 3$, the regime shift is smoothed out as shown in the third row of Fig.~\ref{fig:summary}.

The low average degree of scale free networks, in our case $z = 1.3643$, seems to make the systems less vulnerable.
Also our computations for Poisson random graphs with the same average degree (and thus a parameter $\lambda = 0.6571$, i.e. below the percolation threshold) lead to similar results as for the case of scale free networks.
But, the system size $N$ needs to be considered as well before concluding anything with respect to system safety. 
Our simulations on scale free networks consisting of $N = 1000$ nodes show similar sudden regime shifts as those for Erd\"os-R\'enyi random graphs. 
But, still, the degree distributions of the simulated networks also tend to have higher average degrees than the one used here.

We can verify the simulation results with the help of the cHMF by taking a representative degree distribution of simulated networks as input. 
But, the degree distribution of a simulated network would only converge to the theoretical one used in Equation~(\ref{eq:degrdistr}) in the limit of infinite network size.
In reality though, it might approximate it well for $N = 10^7$. 

Still, we can conclude that in very large systems ($N > 10^7$) where few big hubs coexist with a majority of small degree nodes, the lower connectivity can significantly reduce systemic risk, especially in case of the DD. 

As shown in Fig.~\ref{fig:summary} the DD variant exposes the system to a lower risk than the ED, with a minor exception at the phase transition line for the case of Poisson random graphs.

Still, for both our model variants, a higher risk diversification for every node does not lead to better outcomes in general. 
In fully connected networks - where every node is connected with everyone else and maximal risk diversification is realized - the system is exposed to a higher risk of large cascades in comparison with the studied scale free networks.
The same holds for Poisson random graphs except for threshold parameters close to the phase transition, where the change is so abrupt that we cannot draw any conclusions.

In fact, well diversified - and thus also well connected - nodes have a higher failure risk in regions of high systemic risk in both model variants.

 \begin{figure*}[t]
 \centering
\includegraphics[width=0.3\textwidth]{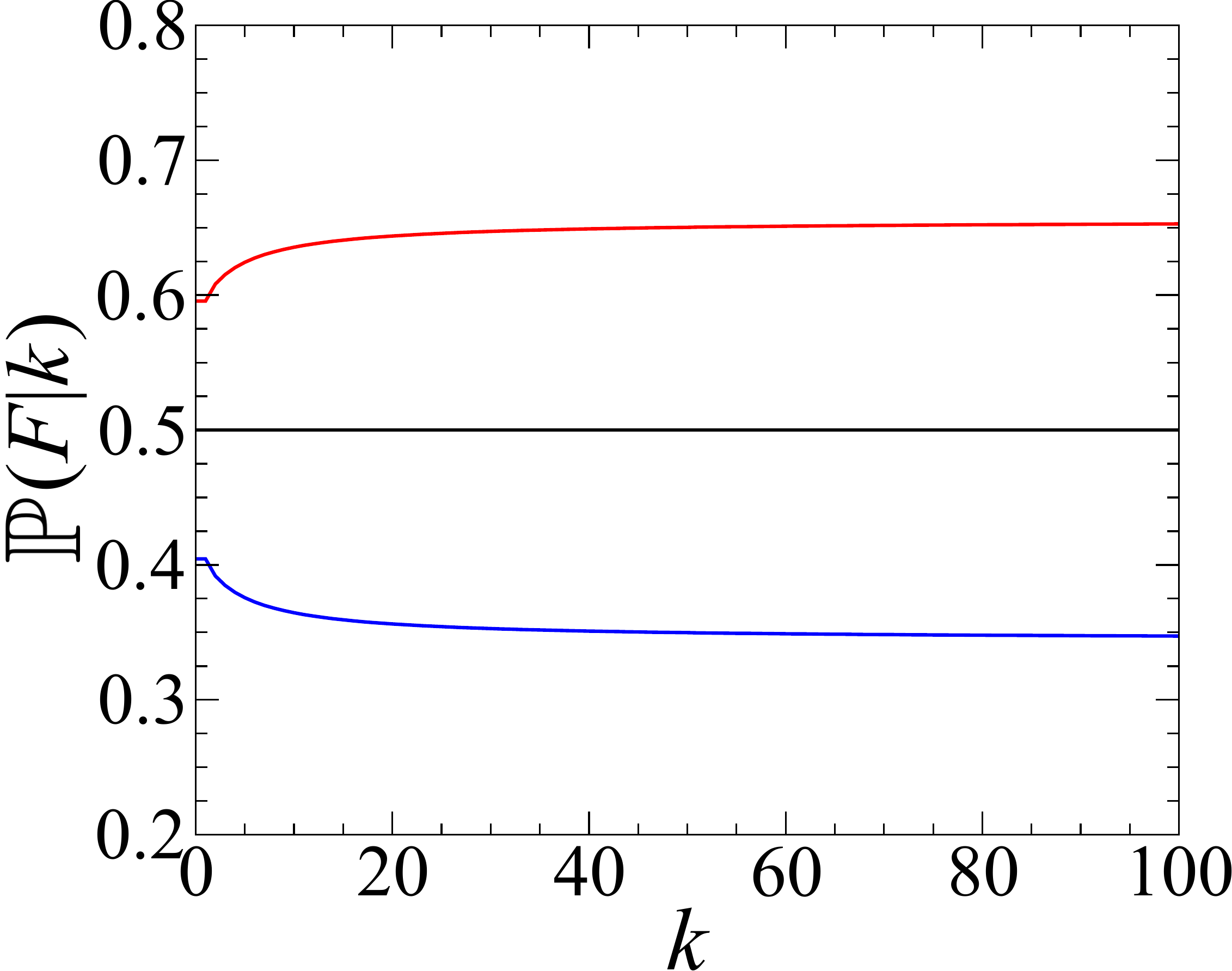}
\includegraphics[width=0.3\textwidth]{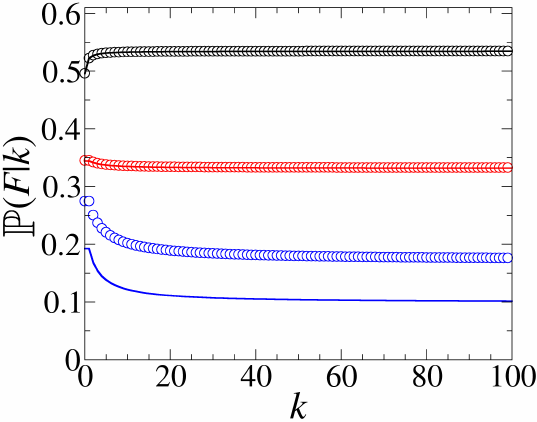}
\includegraphics[width=0.3\textwidth]{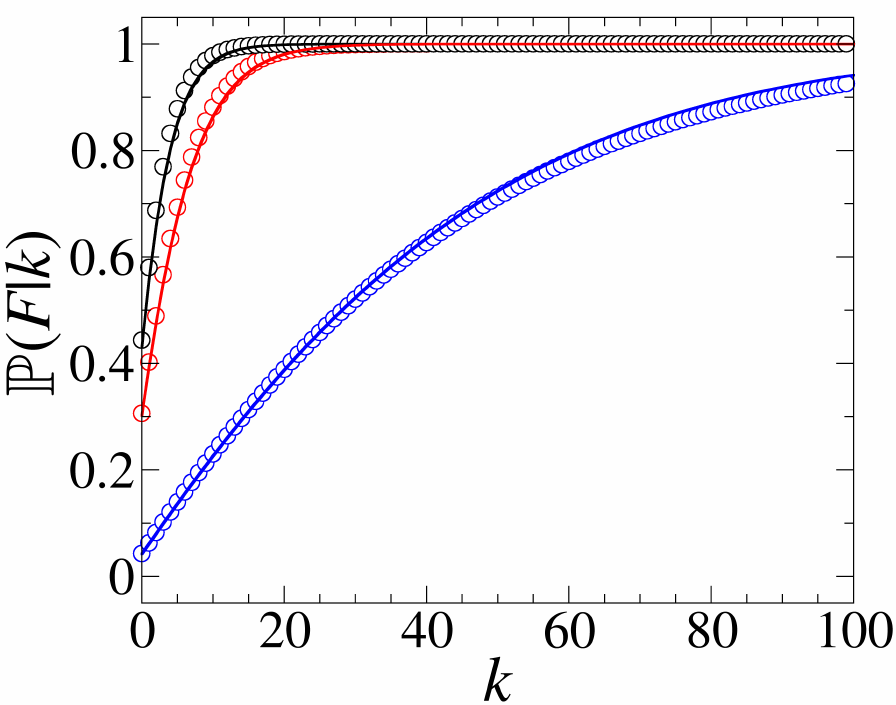}
\begin{picture}(0,0)
    \put(-472,115){(a)}
    \put(-317,115){(b)}
    \put(-155,115){(c)}
   \end{picture}
  \caption{(a) Conditional failure probability for ED and normally distributed thresholds with parameters ${(\mu, \sigma) = (0.5, 0.25)}$ with respect to a neighbor's failure probability of ${\pi = 0.4}$ (blue), ${\pi = 0.5}$ (black) and ${\pi = 0.6}$ (red). (b) Conditional failure probability for ED and normally distributed thresholds with parameters ${(\mu, \sigma) = (0.3, 0.5)}$ (black), ${(\mu, \sigma) = (0.5, 0.6)}$ (red), ${(\mu, \sigma) = (0.5, 0.25)}$ (blue) and $\pi$ defined by cascade equilibrium. Lines indicate a scale free distribution with ${\gamma = 3}$ and ${c = 200}$ and thus ${z = 1.3643}$, while symbols belong to Poisson random graphs with ${\lambda = 0.657}$ and ${c = 100}$ (with the same $z$). (c) As in (b), but for DD.}
\label{fig:pfail}
\end{figure*}
 
\subsection{Individual failure probabilities}

In ED a high diversification is expected to decrease the individual failure risk, since the failure of a high number of neighbors is less probable than the failure of fewer neighbors.
As shown in Fig.~\ref{fig:pfail}(a) and (b), this intuition applies only to regions of lower systemic risk, but tends to saturate for increasing degree $k$ as does the fragility $\phi \simeq 1/k $.
What is not considered in this reasoning is the impact that a failure of a well diversified node has on the overall system stability. Once there is the chance that a few hubs fail, they trigger larger failure cascades in which the failure probability of a neighbor $\pi$ increases. 
Once $\pi$ is big enough, hubs face an even higher failure probability $\mathbb{P}(F\vert k)$ than nodes with smaller degrees as shown in Fig.~\ref{fig:pfail}(a). 
But, also this effect saturates for increasing degrees.

Mathematically, we see that the shape of the individual failure probabilities $\mathbb{P}(F\vert k)$ for ED are defined by only two system sizes: $\pi$, which indicates the overall system state, and the threshold distribution:
\begin{align*}
 \mathbb{P}(F\vert k) = \sum^{k}_{n = 0} b(n, k, \pi) F_{\Theta}\left(\frac{n}{k}\right).
\end{align*}
$\mathbb{P}(F\vert k)$ does not depend directly on the degree distribution, and thus the diversification strategies of the other nodes. 
Still, if $\pi$ exceeds a certain threshold (which is close to $\mu$), nodes with higher degree have a higher failure probability.

In contrast, hubs in DD have always a higher failure probability. Each additional neighbor introduces the possibility of a loss, if it fails. Thus the shape of the conditional failure probability 
\begin{align*}
  \mathbb{P}(F\vert k) = \sum^{k}_{n = 0} b(n, k, \pi)  \sum_{l}p^{*n}_{\rm{imp}}(l) F_{\Theta}(l)
\end{align*}
will always look similar as the ones presented in Fig.~\ref{fig:pfail}(c). 
Consequently, too many nodes with high degrees would increase the vulnerability of the system. 
Still, the presence of hubs also decreases the overall failure risk by decreasing some of the possible losses. 
These losses are distributed by $p^{*n}_{\rm{imp}}(l)$ and can thus reduce all $\mathbb{P}(F\vert k)$. 

Especially less diversified nodes (which have a chance to survive also large failure cascades) profit from the diversification of others.

\section{Discussion}
Although risk diversification is generally considered to lower the risk of an individual node, on the system level it can even lead to the amplification of failures. 
With our work, we have deepened the understanding of cascading failure processes in several ways. 

At first, we generalize the method to calculate the systemic risk measure, i.e. the average cascade size, to include directed and weighted interactions.
This macro measure is complemented by a measure on the meso level, by calculating individual failure probabilities of nodes based on their degree (diversification). 

This allows us to compare two different diversification mechanisms: ED (exposure diversification) and DD (damage diversification). 
As we demonstrate, nodes which diversify their exposures well (i.e. hubs), have a lower failure risk only as long as the system as a whole is relatively robust.
But above a certain threshold for the failure probability of neighboring nodes, such hubs are at higher risk than other nodes because they are more exposed to cascading failures. 
This effect tends to saturate for large degrees.

In general, most regulatory efforts follow the {\it too big to fail} strategy, and focus on the prevention of the failures of systemic relevant nodes - the hubs. 
This is mainly achieved by an increase of the thresholds, i.e., capital buffers in a financial context, but, in reality it could be very costly. 

With our study of another diversification strategy, the damage diversification, we suggest to accompany regulatory efforts by mitigating the failure of hubs.
By limiting the loss that every node can impose on others, the damage potential of hubs and, thus, the overall systemic risk is significantly reduced. 
While this is systemically preferable, the DD strategy is a two-edged sword:  Hubs face an increased failure risk, but many small degree nodes benefit from the diversification of their neighbors.
This lowers the incentives for diversification as long as no other benefit, e.g., higher gains in times of normal system operation, comes along with a high degree.

As we show, the systemic relevance of a node is not solely defined by its degree, or connectivity. 
The size of its impact in case of its failure and thus its ability to cause further failures is crucial.
It is a strength of our approach that we can include this impact analytically and obtain a more refined and realistic identification of system relevant nodes.

If necessary, it is still straight forward to adopt our analytical expansions to the case of directed networks, where the failure of one node would impact another node, but not vice versa.
Additionally, our approach can be transfered to degree-degree correlated networks. 
We would expect that a high degree assortativity in the DD could further reduce systemic risk, since many less diversified nodes could be saved by connections to hubs whose failures would impact their neighborhood only little, but this is outside the scope of the current paper, and will be addressed in a future work.

\paragraph*{Acknowledgements.}
RB acknowledges support by the ETH48 project, RB, AG and FS acknowledge financial support from the Project CR12I1\_127000
{\it OTC Derivatives and Systemic Risk in Financial Networks} financed by the Swiss National Science Foundation.
AG and FS acknowledge support by the EU-FET project MULTIPLEX 317532.

\section{Appendix}

\subsection{Approximation of the convolution}
\label{sec:conv}
For the DD it is necessary but computational expensive to calculate the convolution $p^{*^{n}}_{{\rm{imp}}}$, where \begin{align*}
 p_{\rm{imp}}\left(\frac{1}{k}\right) = \mathbb{P}(F_n\vert k)\frac{k p(k)}{z \pi}
\end{align*}
denotes the probability of a loss $\phi = 1/k$ caused by the failure of one neighbor with degree $k$. 
Given that $n$ neighbors have failed, a node faces a loss $\phi_{n}$, which is a random variable following the law $p^{*^{n}}_{{\rm{imp}}}$.

But the number, $l$, of $\phi_{n}$'s values  with nonzero probability mass, which are of relevant size for good accuracy, as well as the number of accumulation points grows exponentially with the order $n$.

In the end we are interested in calculating the failure probability $\sum_{l}p^{*n}_{\rm{imp}}(l) F_{\Theta}(l)$ of a node so that it suffices to compute $p^{*^{n}}_{{\rm{imp}}}$ accurately on an interval ${[0,b]} \subset \mathbb{R}$, where the threshold distribution $F_{\Theta}$ is effectively smaller than $1$. 
Otherwise we consider the summand  \begin{align*}
 \sum_{l > b} p^{*n}_{\rm{imp}}(l) F_{\Theta}(l) \simeq 1- \sum_{l \leq b} p^{*n}_{\rm{imp}}(l).
\end{align*}
We vary the parameters $\mu$ and $\sigma$ of the threshold distribution between $0$ and $1$ so that we can savely set $b = 5$.
Next, we partition ${[0,b]}$ into $J$ small intervals $I_j = {](j-1)h, j h]}$ of length $h$, with $j = 1, \cdots, J$. 
For small enough $h$ (here $h = 10^{-5}$) it is numerically precise enough to assume the approximation of $p^{*^{n}}_{{\rm{imp}}}$ to be constant on $J$.

\paragraph{Convolution with the help of FFT}
In the standard (and faster) algorithm that we use, we simply bin $p_{{\rm{imp}}}$ to ${[0,b]}$ by
\begin{align*}
 \hat{p}_{{\rm{imp}}}(jh) := \begin{cases}
 \sum_{k: (j-1)h < \frac{1}{k} \leq jh} p(k) & j = 1, \cdots, {J}, \\
 0 & {\rm otherwise}.
 \end{cases}
\end{align*}
Then, we apply the Fast Fourier Transformation (FFT)~\cite{Ruckdeschel2010,Frigo2005}, and take the $n$-th power of the distribution and transform it back in order to receive $\hat{p}^{*^{n}}_{{\rm{imp}}}$.

This is numerically accurate enough for the calculation of the final fraction of failed nodes $\rho^{*}$, but often does not serve well, if we want to deduce the shape of $\mathbb{P}(F\vert k)$. For that purpose we have implemented a more precise alternative.

\paragraph{Alternative convolution algorithm}
Here we do not assume $\hat{p}^{*^{n}}_{{\rm{imp}}}$ to be constant on an interval $I_j$, but uniformly distributed instead. 
For any (discrete) probability distribution $p_X$ we define its approximation in $x \in {[0,b]}$ as
\begin{align*}
 \begin{split}
 a\left(p_{{\rm{imp}}}(x)\right) := & \sum^{J}_{j=1} \frac{x - (j-1)h}{h} \;  \mathbbm{1}_{\left\lbrace (j-1) h < x \leq j h \right\rbrace} \\
 & \sum_{y: (j-1)h < y \leq jh} p_X(y).
 \end{split}
\end{align*}
Thus, we get for $p_{\rm{imp}}$
\begin{align*}
 \begin{split} 
 a\left(p_{{\rm{imp}}}(x)\right) := & \sum^{J}_{j=1} \frac{x - (j-1)h}{h} \;  \mathbbm{1}_{\left\lbrace (j-1) h < x \leq j h \right\rbrace} \\
 & \sum_{k: (j-1)h < \frac{1}{k} \leq jh} \frac{ \mathbb{P}(F_n\vert k) p(k) k}{z \pi}.
 \end{split}
\end{align*}
This is the initial distribution of an iteration in which we compute $\widehat{p^{*^{n}}_{{\rm{imp}}}}$ in the $n$-th step by convoluting it first exactly with $\widehat{p^{*^{n - 1}}_{{\rm{imp}}}}$ of the previous step with the non-approximated $p_{\rm{imp}}$. Afterwards we bin it to the intervals $I_j$ again
\begin{align*}
 \widehat{p^{*^{n}}_{{\rm{imp}}}} := a\left(\widehat{p^{*^{n - 1}}_{{\rm{imp}}}} * p_{\rm{imp}} \right),
\end{align*}
with
\begin{align*}
  \widehat{p^{*^{1}}_{{\rm{imp}}}} :=  a\left(p_{\rm{imp}} \right).
\end{align*}

\bibliographystyle{plain}
\bibliography{BGS_SystRisk_CLU}

\end{document}